\documentclass[journal]{IEEEtran} 

\usepackage[T1]{fontenc}% optional T1 font encoding

\usepackage{amsfonts}
\usepackage{amsthm}
\usepackage{amssymb}
\usepackage{graphicx}
\usepackage{subcaption}
\usepackage{lipsum}
\usepackage{multicol} \usepackage{mathtools}
\usepackage{multirow}
\usepackage{threeparttable}
\usepackage{gensymb}
\usepackage[table,pdftex,hyperref,svgnames]{xcolor} 
\usepackage{color, colortbl}
\definecolor{Gray}{gray}{0.9}

\usepackage{array}
\usepackage[normalem]{ulem}

\usepackage{hyperref}%
\hypersetup{colorlinks=true, linkcolor=blue, breaklinks=true, urlcolor=blue}

\usepackage{multicol} \usepackage{mathtools}
\usepackage{version,xspace}
\usepackage{url,doi}

\newcommand{\subscr}[2]{#1_{\textup{#2}}}

\DeclareMathOperator{\sgn}{sgn}

\newcommand\oprocendsymbol{\hbox{$\triangle$}}
\newcommand\oprocend{\relax\ifmmode\else\unskip\hfill\fi\oprocendsymbol}

\usepackage{enumitem}

\DeclareSymbolFont{bbold}{U}{bbold}{m}{n}
\DeclareSymbolFontAlphabet{\mathbbold}{bbold}
\newcommand{\vect}[1]{\mathbbold{#1}}

\renewcommand{\circ}{\odot}
\newcommand{\scirc}{\raise1pt\hbox{$\,\scriptstyle\circ\,$}}
\newcommand{\real}{\mathbb{R}}

 \newcommand{\cl}{\mathrm{cl}}
\newcommand{\interior}{\mathrm{int}}

\newtheorem{theorem}{Theorem}
\newtheorem{lemma}[theorem]{Lemma}

\newtheorem{example}[theorem]{Example}
\newtheorem{definition}[theorem]{Definition}
\newtheorem{problem}{Problem}

\DeclareMathOperator{\diag}{diag}
\DeclareMathOperator{\spn}{span}

\DeclareMathOperator{\Img}{\mathrm{Img}}
\renewcommand{\top}{\mathsf{T}} %or \top or \intercal

\usepackage{scalerel}
\usepackage{tikz}
\usetikzlibrary{svg.path}
\definecolor{orcidlogocol}{HTML}{A6CE39}
\tikzset{
	orcidlogo/.pic={
		\fill[orcidlogocol] svg{M256,128c0,70.7-57.3,128-128,128C57.3,256,0,198.7,0,128C0,57.3,57.3,0,128,0C198.7,0,256,57.3,256,128z};
		\fill[white] svg{M86.3,186.2H70.9V79.1h15.4v48.4V186.2z}
		svg{M108.9,79.1h41.6c39.6,0,57,28.3,57,53.6c0,27.5-21.5,53.6-56.8,53.6h-41.8V79.1z M124.3,172.4h24.5c34.9,0,42.9-26.5,42.9-39.7c0-21.5-13.7-39.7-43.7-39.7h-23.7V172.4z}
		svg{M88.7,56.8c0,5.5-4.5,10.1-10.1,10.1c-5.6,0-10.1-4.6-10.1-10.1c0-5.6,4.5-10.1,10.1-10.1C84.2,46.7,88.7,51.3,88.7,56.8z};
	}
}
\newcommand\orcidicon[1]{\href{https://orcid.org/#1}{\mbox{\scalerel*{
				\begin{tikzpicture}[yscale=-1,transform shape]
				\pic{orcidlogo};
				\end{tikzpicture}
			}{|}}}}

\newcommand{\rev}[1]{#1}

\begin{document}
	
	\title{Transient Stability of Droop-Controlled Inverter Networks with Operating Constraints}
	
	\author{Kevin~D.~Smith~\textsuperscript{\orcidicon{0000-0002-1407-5893}},~\IEEEmembership{Student Member,~IEEE},
		Saber~Jafarpour~\textsuperscript{\orcidicon{0000-0002-7614-2940}},~\IEEEmembership{Member,~IEEE,}
		and Francesco~Bullo~\textsuperscript{\orcidicon{0000-0002-4785-2118}},~\IEEEmembership{Fellow,~IEEE}%
		
		\IEEEcompsocitemizethanks{%%
			\IEEEcompsocthanksitem This work was supported in part by the
			U.S.\ Department of Energy (DOE) Solar Energy Technologies Office under
			Contract No. DE-EE0000-1583, the National Science Foundation grant
			DGE-1258507, and the U.S.\ Defense Threat Reduction Agency under grant
			HDTRA1-19-1-0017.
			\IEEEcompsocthanksitem Kevin D. Smith, Saber Jafarpour, and Francesco
			Bullo are with the Center
			of Control, Dynamical Systems and Computation, UC Santa Barbara, CA
			93106-5070, USA. {\tt \{kevinsmith,saber.jafarpour,bullo\}@ucsb.edu}}}
	
	\maketitle
	
	\begin{abstract}
		Due to the rise of distributed energy resources, the control of networks
		of grid-forming inverters is now a pressing issue for power system
		operation. Droop control is a popular control strategy in the literature
		for frequency control of these inverters. In this paper, we analyze
		transient stability in droop-controlled inverter networks that are
		subject to multiple operating constraints.  Using a	physically-meaningful 
		Lyapunov-like function, we provide two sets of
		criteria (one mathematical and one computational) to certify that a
		post-fault trajectory achieves frequency synchronization while respecting
		operating constraints. \rev{We show how to obtain less-conservative transient 
		stability conditions by incorporating information from loop flows, i.e., net 
		flows of active power around cycles in the network. Finally, we use these 
		conditions to quantify the scale of parameter disturbances to which the network 
		is robust. We illustrate our results with numerical case studies of the IEEE 
		24-bus system.}
	\end{abstract}
	
	
	\IEEEpeerreviewmaketitle

	\section{Introduction}
	
	Transient stability is a power systems problem of both practical importance and 
	theoretical interest. The goal of transient stability analysis is to determine 
	whether or not the system will return to a stable, frequency-synchronized operating 
	point after a large disturbance. Transients are difficult to analyze: the governing 
	differential equations are nonlinear, and linearization techniques are not useful for 
	large-scale disturbances. Therefore, system operators typically rely on numerical 
	simulation \cite[Chapter 9.3]{LLG:12} to study system behavior. \rev{Simulation 
	is an effective tool for analyzing individual disturbance scenarios, but it has 
	limitations. Simulating a comprehensive set of disturbances is computationally 
	expensive, and it does not establish rigorous guarantees.}
	
	\textit{Direct methods} of transient stability analysis \rev{address these 
	limitations} by establishing theoretical guarantees on transient behavior. 
	\rev{Direct methods are not a substitute for simulation in real-world power system 
	operation, since they rely on 
	low-order, theoretically-tractable models. Instead, they provide significant 
	theoretical insight into these simplified systems.}
	Classical works on using Lyapunov-like methods to study transient stability include 
	\cite{TA-RP-SV:79,PV-FFW-RLC:85,HDC-CCC-GC:95}. More recently, 
	\cite{TLV-KT:16,TLV-KT:17} used set-theoretic control techniques to establish regions 
	of attraction for the coupled swing equations. Direct methods are highly 
	model-specific and provide conservative guarantees, so this topic is still the 
	subject of active research. 
	
	Historically, the literature on direct methods has focused on networks of high-inertia synchronous generators. But the rise of distributed energy resources has sparked a growing interest in the stability of low-inertia inverter networks, particularly microgrids. Inertia is both a blessing and a curse from a control perspective---the same inertia that makes the system robust to disturbances also makes the system respond sluggishly to control inputs. A suitable fast-acting controller can make a low-inertia inverter network highly robust. Two broad classes of inverter controllers have emerged to exploit this low inertia: \textit{grid-following} controllers, in which the inverter acts as a current source to track the local voltage signal; and \textit{grid-forming} controllers, in which the inverter acts as a voltage source to stabilize voltage frequencies throughout the network. Both of these frameworks involve new models and require fresh approaches to direct transient stability analysis.  
	
	One of the most popular approaches to grid-forming control is \textit{proportional 
	droop control}, in which local voltage frequencies are modulated in proportion to the 
	power drawn from neighboring buses. Recent work 
	\cite{JWSP-FD-FB:12u,NA-SG:13,LZ-DH:18} has studied the dynamics of droop-controlled 
	microgrids (DCMGs) \rev{via} the inhomogeneous Kuramoto model. Under certain 
	assumptions, equilibrium points of the Kuramoto model 
	correspond to frequency-synchronized operating points of the DCMG, and regions of 
	attraction around these equilibria provide a rigorous way to assess how robust DCMG 
	operating points are to disturbances. Some progress has been made on \rev{estimating 
	these} regions of attraction \cite{FD-FB:09z}, but these closed-form estimates tend 
	to be very conservative and \rev{require stringent regularity assumptions on the 
	topology or system parameters}. 

	Another limitation of the literature is that few 
	bounds on the transients are available. To a system operator, a guarantee of 
	frequency synchronization alone 
	is not satisfying, if the resulting transient will violate operating constraints 
	(like constraints on line flows and nodal power injections). \rev{Recent work has 
	begun to address transient stability in conjunction with other engineering 
	constraints \cite{SK-SPN-KK-SG-IAH:19, SK-WD-SPN-FT-KS:19}.}
	If \rev{direct methods of transient stability are} to provide more insight into the 
	operation of DCMGs, then 
	less-conservative regions of attraction, as well as bounds on quantities of 
	engineering significance, are needed. This paper addresses these two needs.

	\rev{
	\paragraph*{Contributions} 
	
	Our first contribution is to extend the transient stability problem. In addition to 
	the classical notion of transient stability (asymptotic frequency
	synchronization), we impose five ``desired properties'' of transients, so as to 
	enforce operating constraints on nodal frequencies, power flows across transmission 
	lines, nodal power injections, nodal ramping rates, and reserves of stored energy. 
	
	Our second contribution is to provide two sufficient conditions for when 
	a trajectory of a DCMG will exhibit transient stability and the five 
	desired properties. Both certificates require only two pieces of information from 
	from the initial condition (nodal frequencies and line angle differences) instead 
	of the full (and harder to measure) vector of voltage angles. The first certificate 
	can be viewed as a DCMG-specific form of Nagumo's theorem, and it is intended as a 
	theoretical basis for transient-stability-certifying algorithms. The 
	second certificate, which consists of a tractable mixed-integer linear program 
	(MILP), is built on top of the first certificate. These theoretical results use a 
	physically-meaningful Lyapunov function called the ``maximum frequency deviation,'' 
	which (to our knowledge) has not been used before to study power systems. 
	
	Our third contribution is to improve these two certificates using the winding 
	partition of the $n$-torus. We introduced the winding partition in 
	\cite{SJ-EYH-KDS-FB:18jv3} to localize the multiple equilibrium points of network 
	flows 
	on the $n$-torus. This paper provides the first application of the winding 
	partition to analyzing system dynamics (in contrast to its previous applications to 
	statics problems). We show how to incorporate the ``winding vector'' of the initial 
	condition (a quantity closely related to flows of active power around cycles in the 
	network) into the two certificates, resulting in less-conservative conditions for 
	transient stability and the other desired properties.
	
	As a fourth contribution, we use our transient stability conditions to quantify the 
	size of parameter disturbances with respect to which the DCMG is robust. We define a single number 
	that quantifies the ``size'' of an arbitrary change in model parameters, and we 
	compute a critical threshold such that post-fault transient stability is guaranteed 
	in any disturbance ``smaller'' than the threshold. We examine particular disturbance 
	modes, including changes in nominal power injections, voltage magnitudes, and branch 
	admittance magnitudes. We illustrate all of these results numerically, using the IEEE 
	24-bus system as a case study. 
	
	\paragraph*{Organization}
	
	The next four sections are  organized around our four main contributions. 
	After introducing our notation, model, and problem statement, Section 
	\ref{sect:prelims} states the extended transient stability problem in Definition 
	\ref{def:dp}, formally introducing transient stability and the five desired
	properties of the transient. Section \ref{section:ts} presents our two certificates 
	in Theorem~\ref{thm:general} and Theorem~\ref{thm:milp}, respectively. We 
	review the winding partition in Section \ref{sect:winding} and improve both 
	certificates by incorporating the winding vector in Theorem~\ref{thm:winding}, and we 
	show that these improved certificates are less conservative (Theorem 
	\ref{thm:conservative}). Finally, Section \ref{sect:robust} uses the stability 
	certificates to study how robust a DCMG is to changes in parameters, and it presents 
	our numerical case studies. 
	
	\section{Preliminaries and Problem Statement}
	\label{sect:prelims}
	
	\subsection{Notation}
	
	\paragraph*{The Circle and $n$-Torus}
	Let $\mathbb S$ be the circle, i.e., the set of phases or angles. For every pair of 
	angles $\alpha, \beta \in \mathbb S$, we use $|\alpha - \beta|$ to denote the 
	geodesic distance between them. The counterclockwise difference between two angles is 
	the map $d_{\rm cc}: \mathbb S \times \mathbb S \to [-\pi, \pi)$, where
	\[
		d_{\rm cc}(\alpha, \beta) = \begin{cases}
			|\alpha - \beta|, & \text{c.c. arc from $\alpha$ to $\beta$ shorter than 
				$\pi$} \\
			-|\alpha - \beta|, & \text{otherwise}
		\end{cases}
	\]
	In other words, we consider the clockwise and counterclockwise arcs from $\alpha$ to 
	$\beta$. If the counterclockwise arc is shorter, then $d_{\rm cc}(\alpha, \beta)$ is 
	the length of that arc. Otherwise, $d_{\rm cc}(\alpha, \beta)$ is the negated length 
	of the clockwise arc. The $n$-torus, denoted $\mathbb T^n$, is the product of $n$ 
	circles.
	
	\paragraph*{Sets}
	Given any set $S$ within $\mathbb R^n$ or $\mathbb T^n$, we 
	refer to the interior of the set by $\interior(S)$, the closure by $\cl(S)$, and the 
	boundary by $\partial S = \cl(S) \setminus \interior(S)$, with respect to the 
	standard topologies on $\mathbb R^n$ and $\mathbb T^n$. Given a function $V:\real^n\to
	\real$ and a scalar $c\in \real$, we define a sublevel set
	\[
		V^{-1}_{<}(c) =\{x\in \real^n\mid V(x) < c\}. 
	\]
	
	\paragraph*{Linear Algebra}
	The vector $\vect{1}_n$ (resp. $\vect{0}_n$)
	is a vector in $\real^n$ with all the entries equal to one
	(resp. zero). For every $v\in \real^n$, $\diag(v)\in
	\real^{n\times n}$ is a diagonal matrix with entries $\diag(v)_{ii} = v_i$
	for every $i\in \{1,\ldots,n\}$. The $\infty$-norm of $v$ is
	$\|v\|_{\infty} = \max_i|v_i|$, and the $1$-norm of $v$ is
	$\|v\|_1=\sum_{i=1}^{n}|v_i|$. We define $v_{\rm sum} = \sum_{i=1}^n
	v_i$ and $\rev{v_{\rm min}} = \min_i \{v_i\}$. For every $v,w\in \real^n$,
	we write $v \le w$ (resp. $v < w$) if $v_i \le w_i$ (resp. $v_i < w_i$), 
	for every $i\in \{1,\ldots,n\}$. For a matrix $X\in \real^{n\times n}$, the 
	Moore\textendash{}Penrose pseudoinverse
	is denoted by $X^{\dagger}$.  
	
	\paragraph*{Graph Theory}
	An undirected graph is a pair $G = (\mathcal V, \mathcal E)$, where $\mathcal V$ is a 
	set of $n$ nodes, and $\mathcal E \subseteq \mathcal V \times \mathcal V$ is the set 
	of $m$ edges. The neighborhood of any node $i \in \mathcal V$ is denoted by 
	$\mathcal 
	  N(i)$. While $G$ is undirected, we may enumerate and assign an arbitrary orientation to each
          edge $e \in 
	\mathcal E$ by labeling one incident node as the ``source'' $s(e)$ and the other as 
	the ``sink'' $t(e)$. 
	The incidence matrix of the graph \cite[\S 9.1]{FB:20} is the matrix $B \in \{-1, 0, 
	1\}^{n \times m}$ with entries
	\[
		B_{i,e} = \begin{cases}
			+1, & s(e) = i \\
			-1, & t(e) = i \\
			0, & \text{otherwise}
		\end{cases}
	\]
	
	\paragraph*{Graphs and the $n$-Torus}
	We may assign a phase-valued state to every node in $G$, so that the full state of 
	the graph is in $\mathbb T^n$. Given a state $\theta \in \mathbb T^n$, we use the 
	abuse of notation $B^\top \theta$ to represent the vector in $\mathbb R^m$ of 
	counterclockwise differences across each edge, i.e., $(B^\top \theta)_e = d_{\rm 
	cc}(\theta_i, \theta_j)$, where $i$ is the source of $e$ and $j$ is the sink. 
	Furthermore, given any $\gamma \in (0, \pi]^m$, we define the phase-cohesive 
	set as the open set
	\[
		\Delta(\gamma) = \left\{
			\theta \in \mathbb T^n : |B^\top \theta| < \gamma
		\right\} 
	\]
	In contrast to the literature, where $\Delta(\gamma)$ takes a scalar-valued 
	$\gamma$, the $\gamma$ we refer to in this paper is always a vector, allowing for 
	inhomogeneous phase cohesion.
}
	
\rev{
	\subsection{Model}
	
	We consider a DCMG on an undirected topology $G = (\mathcal V, \mathcal E)$, with $n$ 
	buses $\mathcal V = \{1, \dots, n\}$ and $m$ branches (or lines) $\mathcal E 
	\subseteq \mathcal V 
	\times \mathcal V$. We assume that $G$ is connected, but otherwise we make no 
	assumptions about its structure; both trees and cyclic graphs are acceptable. 
	
	\paragraph*{Bus Model} 
	Each bus has a complex voltage $E_i e^{j \theta_i}$, where $E_i > 0$ is the voltage 
	magnitude and $\theta_i \in \mathbb S$ is the phase. We assume that voltage 
	controllers are operating at a much faster time scale than frequency controllers, so 
	that $E_i$ is constant but $\theta_i$ is dynamic. We consider two types of buses: 
	droop-controlled inverters and frequency-dependent 
	loads. Buses in $\mathcal V_I \subset \mathcal V$ represent droop-controlled 
	inverters, which produce a controllable voltage signal with constant magnitude $E_i$ 
	and time-varying frequency $\dot \theta_i$. These inverters operate according to the 
	frequency droop control law \cite{MCC-DMD-RA:93, JMG-JCV-JM-LGDV-MC:11}:
	\begin{equation}
	\dot \theta_i = \omega^* - \frac{p_{e, i} - p_i^*}{d_i}, \qquad \forall i \in 
	\mathcal V_I
	\label{eq:droop}
	\end{equation}
	Here $\dot \theta_i$ is the instantaneous AC frequency, $\omega^*$ is the nominal 
	frequency (for example, 60 Hz), $p_i^* \ge 0$ is the nominal active power injection, 
	and $d_i^{-1} > 0$ is the droop coefficient. Buses in $\mathcal V_L = \mathcal V 
	\setminus \mathcal V_I$ represent frequency-dependent loads \cite[\S 9.1]{PK:94}, 
	where the instantaneous active power injection $p_{e,i}$ is
	\begin{equation}
	p_{e,i} = p_i^* - d_i  (\dot \theta_i - \omega^*), \qquad \forall i \in \mathcal V_L
	\label{eq:load}
	\end{equation}
	Here $p_i^* \le 0$ is the nominal active power load, and $d_i > 0$. Note that 
	\eqref{eq:droop} and \eqref{eq:load} are algebraically equivalent.
	
	\paragraph*{Branch Model} 
	For each branch $\{i, j\} \in \mathcal E$, we assume that the real power flow from 
	node $i$ into the $\{i,j\}$ branch is
	\begin{equation}
		p_{ij}^{\rm line} = \tilde a_{ij} + a_{ij} \sin(\theta_i - \theta_j - 
		\phi_{ij})
		\label{p:line}
	\end{equation}
	where $\tilde a_{ij} \in \real$, $a_{ij} \ge 0$ and $\phi_{ij} \in (-\frac \pi 2, 
	\frac \pi 2)$ are constants. These constants are not necessarily symmetric (i.e., 
	$\phi_{ij} \ne \phi_{ji}$), so in general $p_{ij} \ne -p_{ji}$. 
	
	The AC steady-state active power flow across many types of branches 
	can be written in the form \eqref{p:line}. Transmission lines, for example, are 
	typically
	represented by the nominal $\Pi$ model, which consists of a series admittance $Y_{ij} 
	e^{j \varphi_{ij}}$ that is flanked by two shunt admittances $Y_{ii} e^{j 
	\varphi_{ii}}$ and $Y_{jj} e^{j \varphi_{jj}}$ \cite[\S 6.1]{PK:94}. Active power 
	flow in the nominal $\Pi$ model is given by \eqref{p:line} with $\tilde a_{ij} = 
	E_i^2 (Y_{ii} \cos(\varphi_{ii}) + Y_{ij} \cos(\varphi_{ij}))$, $a_{ij} = E_i E_j 
	Y_{ij}$, and $\phi_{ij} = \varphi_{ij} + \frac \pi 2$. In medium-length and 
	short-length lines, the shunt admittance is typically purely capacitive or altogether 
	negligible, either of which leads to the simplification $\tilde a_{ij} = E_i^2 Y_{ij} 
	\cos(\varphi_{ij})$. It is also typical that the series admittance is primarily 
	inductive, so $\phi_{ij} \approx 0$. In the extreme case of lossless lines (with no 
	shunt admittance), the active power flow reduces to the antisymmetric form $p_{ij}^{\rm 
	line} = a_{ij} \sin(\theta_i - \theta_j)$. For transformers, active power flow in the 
	standard equivalent circuit model can also be written in the form 
	\eqref{p:line}. We omit the specifications of the parameters for brevity and 
	refer the reader to \cite[\S 6.2]{PK:94}. 
	
	\paragraph*{Dynamics}
	
	Due to conservation of energy, active power injections at 
	each bus must balance against power outflows:
	\[
		p_{e,i} = \sum_{j \in \mathcal N(i)} \tilde a_{ij} + a_{ij} \sin(\theta_i - 
		\theta_j - \phi_{ij}), \qquad \forall i \in \mathcal V
	\]
	Substituting this expression for $p_{e,i}$ into \eqref{eq:droop} and \eqref{eq:load} 
	leads to a differential equation in $\theta$ that captures the angle dynamics of the 
	grid. We can write these dynamics compactly by defining a constant vector $p 
	\in \mathbb R^{n}$ with entries $p_i = p_i^* + \omega^* d_i - \sum_{j \in \mathcal 
	N(i)} \tilde a_{ij}$ for each $i \in \mathcal V$, as well as a matrix $D = 
	\diag\{d_i, \; i \in \mathcal V\}$. Then the system can be written as
	\begin{equation}
		D \dot \theta = f(\theta) \label{eq:dynamics}
	\end{equation}
	where $f: \mathbb T^n \to \mathbb R^n$ is a vector with entries
	\[
		f_i(\theta) = p_i - \sum_{j \in \mathcal N(i)} a_{ij} 
		\sin(\theta_i - \theta_j - \phi_{ij}), \qquad \forall i \in \mathcal V
	\] 
	Equation \eqref{eq:dynamics} is the model that we study in this paper. If the sine 
	coefficients are homogeneous and the underlying graph is complete, this model is 
	familiar in the physics community as the Kuramoto-Sakaguchi model 
	\cite{HS-YK:86}, which has been used to study synchronization phenomena in coupled 
	oscillator networks \cite{MKSY-SHS:99,FDS-DA:07,JCB-TC-LD:18}.
	
	\paragraph*{Limitations of the Model}
	
	Our model is based on several commonly-used simplifying assumptions that should be 
	examined explicitly. Perhaps the most important simplification is that we 
	neglect voltage dynamics and reactive power. This is particularly common in the 
	controls community, and it is 
	often justified by assuming fast-acting voltage controllers \cite{TLV-KT:17, 
	KX-HXL-CS-JHVS:20}. If voltage control fails, possibly due to insufficient reactive 
	power, then unmodeled dynamics of the $a_{ij}$ and $\tilde a_{ij}$ parameters may 
	destabilize the system.
	
	Another simplification is our use of steady-state AC models for branches in 
	\eqref{p:line}, which is very common in analysis of conventional power grids. These 
	models assume sinusoidal nodal voltages at a constant 
	frequency, an assumption that is technically contradicted by the dynamic frequencies 
	in \eqref{eq:dynamics}. But the purpose of this paper is to find sufficient 
	conditions for ``safe'' transient stability, and a key aspect of safe power grid 
	operation is a tight tolerance around the nominal frequency, typically under 1\%. In 
	other words, the trajectories that we are interested in certifying have only a small 
	variance in frequency. Nonetheless, the effects of transmission line dynamics on 
	inverter-based grids is a subject of recent interest, and we refer the reader to 
	\cite{DG-JSB-FD:19} for a rigorous study of this topic.
}	
	
	\subsection{Problem Statement}
	Under normal operation, nodal frequencies are synchronized at the nominal frequency 
	$\omega^*$, and power injections $p_{e}$ are equal to the nominal power injections 
	$p^*$. But contingencies, like failing transmission lines or a sudden change in power 
	supply or demand, disrupt this equilibrium behavior. Droop control will stabilize the 
	post-fault system about a new equilibrium, provided that this new equilibrium is 
	sufficiently close to the pre-fault state. This local stability property is 
	well-known and easily verified by eigenvalue analysis of \eqref{eq:dynamics}. 
	
	Unfortunately, the dynamics of droop control after larger-scale
	disturbances are not as well understood, and local stability alone does not inspire 
	confidence in a power system controller. \rev{Furthermore}, the 
	controller should ensure that the system's critical engineering constraints are 
	satisfied \rev{during the transient}. In this paper, \rev{in addition to non-local 
	transient stability, we consider five engineering} constraints that are important 
	in the context of inverter networks:
	
	\begin{definition}[Desired Properties]
		\label{def:dp}
		We define the following six properties that are desirable in a trajectory 
		$\theta(t)$ of \eqref{eq:dynamics}:
		\begin{enumerate}[label=\textup{(P\arabic*)}]
			\item \label{dp:stability}
			\emph{Transient stability.} Nodal frequencies asymptotically synchronize, i.e., $\lim_{t \to \infty} \dot \theta(t) = \subscr \omega {syn} \vect 1_n$ for some synchronous frequency $\subscr \omega {syn} \in \mathbb R$.
			\item \label{dp:frequency}
			\emph{Frequency constraint.} Nodal frequencies are
			bounded by $|\dot \theta(t) - \omega^*
			\vect 1_n| \le \bar \delta$ for all $t \ge 0$, where
			$\bar \delta \ge \vect 0_n$ is a vector of frequency
			tolerances.
			\item \label{dp:angle}
			\emph{Angle difference constraint.} Voltage angle
			differences are bounded by $|B^\top \theta(t)| \le
			\bar \gamma$ for all $t \ge 0$, where $\bar \gamma \in
			(0, \frac \pi 2]^m$ is a vector of angle difference
			tolerances.
			\item \label{dp:power}
			\emph{Power constraint.} Power injections are sufficiently close to the nominal injection, i.e.,  $|p_e(t) - p^*| \le \bar p_e$ for all $t \ge 0$, where $\bar p_e \in \mathbb R_{\ge 0}^n$ is a vector of power tolerances.
			\item \label{dp:ramping}
			\emph{Ramping constraint.} The rate of change in power injections is sufficiently small: $|\dot p_e(t)| \le \bar r_e$ for all $t \ge 0$, where $\bar r_e \in \mathbb R_{\ge 0}^n$ is a vector of ramping tolerances.
			\item \label{dp:energy}
			\emph{Energy constraint.} The difference from nominal energy injection is bounded by
			\[
			\left| \int^\infty_0 p_e(t) - p^*~dt \right| \le \bar s
			\]
			where $\bar s \in \mathbb R_{\ge 0}^n$ is a vector of nodal capacities to store or dump energy. 
		\end{enumerate}
  \end{definition}

	\noindent
	\rev{
	Each of these desired properties are necessary for safe operation of the power 
	system. \ref{dp:stability} and \ref{dp:frequency} are the standard objectives of 
	primary and secondary frequency control, which keep nodal frequencies close to the 
	rated frequency of grid components. \ref{dp:angle} protects transmission lines from 
	overheating, since larger angle differences lead to larger current flow, and thus, 
	more thermal dissipation. \ref{dp:power} and \ref{dp:energy} ensure that the power 
	and energy drawn from inverters are within a reasonable range. For example, an 
	inverter powered by solar panels on a sunny afternoon is more flexible in its active 
	power injection (via curtailment) than the same inverter on a cloudy morning. 
	Finally, \ref{dp:ramping} ensures that the rate at which power injections fluctuate 
	is within the tolerance of the inverter.  
}
  The objective of this paper is to find
  computationally-tractable sufficient conditions on
  $\theta(0)$ for each of these six properties.

	\section{Main Theoretical Results}
	\label{section:ts}
	
	\rev{We now proceed with our main results: two sets of sufficient conditions to 
	certify that a trajectory  satisfies transient stability and the five desired 
	properties in Definition \ref{def:dp}.}

	\subsection{Lyapunov Function}
	\label{sect:lyap}
	
	Our analysis is based on the \emph{frequency deviation vector}, which measures the 
	difference between instantaneous and nominal frequencies at each bus:
	\begin{equation}
	\label{def:freq-deviation}
	v(\theta) = \dot \theta - \omega^* \vect 1_n = D^{-1} f(\theta) - \omega^* \vect 1_n
	\end{equation}
	From $v(\theta)$, we define \rev{our} Lyapunov candidate function, the 
	\textit{maximum frequency deviation} 
	\[
		V(\theta) = ||v(\theta)||_\infty
	\]
	If a trajectory $\theta(t)$ is clear from context, we will abuse notation and write 
	$V(t)$ instead of $V(\theta(t))$. \rev{We will show that the min-max frequency 
	deviation is non-increasing when voltage angle differences are sufficiently small;
	exactly how small depends on $\phi_{ij}$. For each branch, we define a 
	\textit{critical arc length}
	\[
		\gamma^*_e = \frac \pi 2 - \max\left\{ |\phi_{ij}|, |\phi_{ji}| \right\}, 
		\qquad \forall e = \{i, j\} \in 
		\mathcal E
	\]
	Due to the assumption that $\phi_{ij} \in (-\frac \pi 2, \frac \pi 2)$, the critical arc 
	lengths satisfy the bound $\gamma_{e}^* \in (0, \frac \pi 2]$, and the maximum 
	value of $\frac \pi 2$ is achieved by lossless transmission lines (for which 
	$\phi_{ij} = 0$). We collect the critical arc lengths into a vector 
	$\gamma^* \in \real^m$.
	
	As long as the angle difference across each branch is less than the critical arc 
	length, the maximum frequency deviation is non-increasing: 
	\begin{lemma}[Max Frequency Deviation is Non-Increasing] \label{thm:lyap}
		Let $\theta(t)$ be a trajectory of \eqref{eq:dynamics} such that $\theta(t) \in 
		\Delta(\gamma^*)$ on some interval $t \in [t_0, t_1]$. Then $V(t_1) \le 
		V(t_0)$.
	\end{lemma}

	\noindent
	The proof of this property is based on the following lemma, which is (to our 
	knowledge) novel:
}

	\begin{lemma}[Sign-Definiteness of Laplacian Matrices]\label{lem:sgnlemma}
		Let \rev{$L \in \mathbb R^{n \times n}$ be a Laplacian matrix.} For any $x \in 
		\mathbb 
		R^n$, let $I_{\rm max} = \{i : |x_i| = ||x||_\infty\}$ be the set of nodes with 
		maximal absolute value. Then
		\[
			\max_{i \in I_{\rm max}} \{ -\sgn(x_i) (L x)_i \} \le 0
		\]
		Furthermore, if the digraph corresponding to $L$ is strongly connected, 
		then equality holds if and only if $x \in \spn(\vect 1_n)$. 
	\end{lemma} 

\rev{
	
	\begin{proof}[Proof of Lemma \ref{lem:sgnlemma}]
		For every $i\in I_{\rm max}$, we compute
		\begin{align*}
		-\sgn(x_i) (L x)_i &= \sum_{j\ne i} L_{ij}|x_i| -
		\sum_{j\ne i} L_{ij} \sgn(x_i) x_j \\
		&\le \sum_{j \ne i} L_{ij} ( |x_i| -|x_j|)
		\end{align*}
		The first line follows because $L$ has zero row sums and the 
		second line because off-diagonal entries of $L$ are non-positive. 
		But $i\in I_{\rm max}$ implies that $|x_i| - |x_j| \ge 0$, so we conclude that 
		$-\sgn(x_i) (Lx)_i \le 0$. Equality clearly holds in the case where $x \in 
		\spn(\vect 1_n)$. Now suppose that $\max_{i \in I_{\rm max}} \{-\sgn(x_i) 
		(Lx)_i\} = 0$, which implies that $\sum_{j \ne i} L_{ij} (|x_i| - \sgn(x_i) x_j) 
		= 0$ for some particular $i \in I_{\rm max}$. But each summand is non-positive, 
		so $|x_i| = \sgn(x_i) x_j$ for each $j$ for which $L_{ij} \ne 0$; consequently, 
		$x_i = x_j$ for each out-neighbor $j$ of $i$. It follows that $j \in I_{\rm 
			max}$. Extending the same argument to $j$ and all of its neighbors, we see 
			that 
		if a directed path exists from $i$ to any node $k$, then $x_k = x_i$. But the 
		graph is strongly connected, so we conclude that $x \in \spn(\vect 1_n)$. 
	\end{proof}  

	\begin{proof}[Proof of Lemma \ref{thm:lyap}]
		We first observe that $\ddot \theta = D^{-1} J(\theta) \dot \theta$, where  
		$J(\theta)$ is the Jacobian matrix of $f(\theta)$. For $i \ne j$,
		\[
			J_{ij}(\theta) = \frac{\partial f_{i}(\theta)}{\partial \theta_j} = 
			\begin{cases}
				-\cos(\theta_i - \theta_j - \phi_{ij}), & \{i, j\} \in \mathcal E \\
				0, & \text{else}
			\end{cases}
		\]
		If $\theta \in \Delta(\gamma^*)$, then $\cos(\theta_i - \theta_j - \phi_{ij}) > 
		0$. Furthermore, evaluating 
		the diagonal entries of $J(\theta)$ reveals that the matrix has zero row sums, so 
		$-J(\theta)$ is the Laplacian matrix of a weighted, directed graph whose topology 
		is identical to $G$ (treating each undirected edge in $G$ is a pair of directed 
		edges). Note that this graph is strongly connected. 
		
		Let $I_{\rm max} = \{i : |v_i(t)| = ||v(t)||_\infty \}$ be the set of buses with 
		maximal frequency deviation, so we can write
		\[
			V(t) = \max_{i \in I_{\rm max}} \left\{ \sgn(v_i(t)) v_i(t) \right\}
		\]
		Using \cite[Lemma 15.16(iii)]{FB:20} to compute the upper right Dini derivative 
		of a pointwise-maximum function, we obtain 
		\begin{align*}
			\text D^+ V(t)
			&= \max_{i \in I_{\rm max}} \left\{ \sgn(v_i(t)) \left( 
			D^{-1} J(\theta) 
			\dot \theta \right)_i  \right\} \\ 
			&= \max_{i \in I_{\rm max}} \left\{ \sgn(v_i(t)) \left( D^{-1} J(\theta) v(t) 
			\right)_i \right\}
		\end{align*}
		where the last step follows because $\omega^* \vect 1_n \in \ker(J)$. But $D^{-1} 
		J(\theta)$ is a Laplacian matrix corresponding to a strongly connected graph, so 
		by Lemma \ref{lem:sgnlemma}, $D^+ V(t) \le 0$. Lemma \ref{thm:lyap} 
		follows from this bound \cite[Lemma 15.16(ii)]{FB:20}. 
	\end{proof} 

	In summary, the maximum frequency deviation is positive definite about the subspace 
	of frequency-synchronized states, and it is non-increasing inside of 
	$\Delta(\gamma^*)$.  
}
	
	\subsection{Set-Theoretic Certificate}
	\label{sect:general}

	We now use \rev{the maximum frequency deviation to} establish a set-theoretic 
	transient stability and operating constraint certification. Our approach is to 
	construct forward-invariant sets using $V$, \rev{based on} the following 
	optimization problem:
	
	\begin{problem}[Min-Max Frequency Deviation]
		Let $S \subseteq \Delta(\gamma^*)$. We define $V^*(\partial S)$ to be the 
		minimum value of the following:
		\[
		\begin{array}{rl}
		\text{minimize}: & V(\theta) \\
		\text{variables}: & \theta \in \mathbb T^n \\
		\text{subject to}: & \theta \in \partial S \\
		& D^{-1} f(\theta) ~\text{is pointed outward from}~S
		\end{array} 
		\]
		If the problem is infeasible, we define $V^*(\partial S) = +\infty$.
		\label{prob:minmax}
	\end{problem}
	
	\noindent
	The min-max frequency deviation is the minimum value of $V(\theta)$ along the 
	``outward boundary'' of $S$, i.e., the portion of $\partial S$ where $\dot \theta$ is 
	pointed away from the set. 
	
	Minimizing a Lyapunov function around a set boundary is a well-established technique 
	for constructing forward-invariant sets---see, for example, Nagumo's 1942 theorem 
	\cite[Theorem 4.7]{FB-SM:15}. More recently, \cite{TLV-KT:17} applied this technique 
	to a quadratic Lyapunov function for the coupled swing equations. In our case, the 
	min-max frequency deviation is defined so that sets of the form $S \cap 
	V^{-1}_{<}(V^*(\partial S))$ are forward invariant. This observation, 
	together with the \rev{monotonicity of $V$, leads} to the central theorem of the 
	paper.
	
	\begin{theorem}[Set-Theoretic Certificate] \label{thm:general}
		Let $\theta(t)$ be a trajectory of \eqref{eq:dynamics}. Let
		$\gamma_0 = |B^\top \theta(0)|$ and $\delta_0 = V(\theta(0))$
		denote the initial angle differences and max frequency deviation.
		\rev{If there exist a vector $\gamma \in 
		[\gamma_0, \gamma^*]$ and a set $\Delta(\gamma_0) \subseteq S \subseteq 
		\rev{\Delta(\gamma)}$ such that $\delta_0 < V^*(\partial S)$}, then
		\begin{enumerate}
			\item \label{p:fi} $\theta(t) \in S \cap V_{<}^{-1}(\delta_0)$ for all 
			$t \ge 0$.
			\item \label{p:stability}
			The transient stability property \ref{dp:stability} is satisfied.
		\end{enumerate}
		Further conditions on $\gamma$ and $\delta_0$ lead to various
		desirable properties from Definition~\ref{def:dp}:
		\begin{enumerate}
			\setcounter{enumi}{2}
			\item \label{p:freq}
			The frequency constraint \ref{dp:frequency} is satisfied for each bus $i$ if $\delta_0 \le \bar \delta_i$.
			\item \label{p:angle}
			The angle difference constraint \ref{dp:angle} is satisfied for each line $\{i, j\}$ if $\gamma_{ij} \le \bar \gamma_{ij}$.
			\item \label{p:power}
			The power constraint \ref{dp:power} is satisfied for each bus $i$ if $\delta_0 \le \bar p_{e,i} d_i^{-1}$.
			\item \label{p:ramping}
			The ramping constraint \ref{dp:ramping} is satisfied for each bus $i$ if 
			\[
				\delta_0 \le \frac 1 2 \bar r_i \left( \sum_{j \in \mathcal N(i)} a_{ij} 
				\right)^{-1}
			\]
		\end{enumerate}
		\rev{
		Additionally, in the special case of lossless networks (where $a_{ij} = a_{ji}$ 
		and $\phi_{ij} = \phi_{ji} = 0$), the following is true:}
		\begin{enumerate}
			\setcounter{enumi}{6}
			\item \label{p:energy}
			The energy constraint \ref{dp:energy} is satisfied for each bus $i$ if 
			\[
			\delta_0 \le \frac{\lambda_2(L) \cos(\gamma_{\rm max}) \bar s_i}{d_i / d_{\rm 
			min}} \left( 
			1 + \frac 1 2 \log\left( \frac{d_{\rm sum}}{d_{\rm min}} \right) \right)^{-1}
			\]
			where $\lambda_2(L)$ is the smallest non-zero eigenvalue of the Laplacian 
			matrix $L = B \left( \mathcal \diag\{ a_{ij}\}_{\{i, j\} \in \mathcal E} 
			\right) B^\top$.
		\end{enumerate}
	\end{theorem}
 
	\begin{proof}
		To prove statement~\ref{p:fi}, \rev{observe that} any trajectory which escapes 
		$S$ must cross through some point on $\partial S$ where $\dot \theta$ is pointed 
		outward from $S$. By definition, 
		$V^*(\partial S) \le V(\theta)$ at such a point $\theta$. But 
		Lemma \ref{thm:lyap} implies that $V(\theta) \le 
		V(0)$, which further implies that $\theta(0) \notin V_{<}^{-1}(\delta_0)$ \rev{if 
		the trajectory reaches this point.} Forward invariance of $S \cap 
		V_{<}^{-1}(\delta_0)$ follows by contrapositive. \rev{Regarding 
		statement~\ref{p:stability}, recall from the proof of Lemma 
       \ref{thm:lyap} that the frequency dynamics can be written $\ddot \theta = D^{-1} 
       J(\theta) \dot \theta$, where $D^{-1} J(\theta)$ is the negated Laplacian matrix 
       of a strongly connected digraph when $\theta \in \Delta(\gamma^*)$. It follows 
       from~\cite[Theorem~12.10]{FB:20}  that $\dot \theta(t)$ converges to a consensus state.}

                %% \cite[Theorem 1]{LM:04}
                
   		\rev{Statements \ref{p:freq} and \ref{p:angle} follow trivially from statement 
   		\ref{p:fi}.} Statement \ref{p:power} follows because droop control relates power 
   		injections to 
		frequencies by $p_{e,i} = p_i^* - d_i( \dot \theta_i - \omega^*)$ for each $i \in 
		\mathcal V$. Therefore $|p_{e,i}(t) - p_i^*| \le d_i V(t) \le d_i \delta_0$ for 
		all $t \ge 0$. To prove statement~\ref{p:ramping}, we observe for each bus $i$ 
		that
		\[
		|\dot p_{e,i}| = \left| \sum_{j \in \mathcal N(i)} a_{ij} \cos(\theta_i - 
		\theta_j - \phi_{ij}) (\dot \theta_i - \dot \theta_j) \right|
		\le 2 \delta_0 \sum_{j \in \mathcal N(i)} a_{ij} 
		\]
		since $\cos(\theta_i - \theta_j - \phi_{ij}) \in (0, 1)$. \rev{
			To prove \ref{p:energy}, observe that 
			\[
				\frac{d}{dt} v(\theta)^\top D v(\theta) = 2v(\theta)^\top J(\theta) \dot 
				\theta = 2 v(\theta)^\top J(\theta) v(\theta)
			\]
			where the last step follows because $\ker(J(\theta)) = \spn\{\vect 1_n\}$. 
			Under the lossless assumption, $J(\theta)$ is negated symmetric Laplacian 
			matrix, and the edge weights in the corresponding graph are $a_{ij} 
			\cos(\theta_i - \theta_j)$, which is lower-bounded by $a_{ij} 
			\cos(\gamma_{\rm max})$. It follows from 
			\cite[Lemma 6.9(ii)]{FB:20} that $\lambda(-J(\theta)) \ge \cos(\gamma_{\rm 
			max}) \lambda_2(L)$, so
			\begin{align*}
				\frac{d}{dt} v(\theta)^\top D v(\theta) &\le -\cos(\gamma_{\rm max}) 
				\lambda_2(L) ||v(\theta)||_2^2 \\
				&\le -\cos(\gamma_{\rm max}) \lambda_2(L) d_{\rm min} \left( 
				v(\theta)^\top D v(\theta) \right)
			\end{align*}
			Therefore $v(\theta)^\top D v(\theta)$ has an exponential upper bound, which 
			decays in time at the rate $\cos(\gamma_{\rm max}) \lambda_2(L) d_{\rm 
			min}$. The integrand in \ref{dp:energy} can be upper-bounded using both 
			$\delta_0$ and this exponential, yielding the condition in statement 
			\ref{p:energy}.}
	\end{proof}

	\noindent
	
	Theorem~\ref{thm:general} simplifies transient analysis in two ways. First, the 
	conditions depend on the quantities $\gamma_0$ and $\delta_0$, rather than the full 
	initial state $\theta(0)$. A system operator can measure $\gamma_0$ through line 
	flows and $\delta_0$ through nodal frequencies, rather than using state estimation to 
	obtain $\theta(0)$. Second, the theorem recasts transient analysis as the search for 
	a set $S \subseteq \rev{\Delta(\gamma^*)}$ with a sufficiently large min-max 
	frequency deviation. The remainder of the section examines a 
	computationally-efficient way to search for such a set.
	
	\subsection{Groundwork for the MILP Certificate}
	
	It is impractical to repeatedly evaluate $V^*(\partial S)$ while 
	searching for a set that satisfies Theorem~\ref{thm:general}. Fortunately, we can 
	efficiently compute upper bounds on $V^*(\partial S)$ if we restrict our 
	search to sets of the form $S = \Delta(\gamma)$. We obtain these upper 
	bounds through a series of relaxations to Problem \ref{prob:minmax}, and then we
	use these bounds to establish an easily-computable transient stability certificate. 
\rev{This subsection lays out the first of two relaxations that we make for this 
	certificate.
	
	When $S = \Delta(\gamma)$ for some $\gamma \in (0, \gamma^*]$, Problem 
	\ref{prob:minmax} admits the following relaxation:
	
	\begin{problem}[Min-Max Frequency Deviation, Lower Bound]
		Let $\gamma \in (0, \gamma^*]$. We define $\widehat V(\gamma)$ to be 
		the minimum value of the following:
		\begin{subequations}
		\begin{align}  
			&\!\min & & ||D^{-1} f - \omega^* \vect 1_n||_{\infty} \label{prob2:cost} 
			\\
			&\text{w.r.t.} & & f \in \mathbb R^n, \; y \in \mathbb R^m, \; \eta^+ \in 
			\mathbb R^m, \; \eta^- \in \mathbb R^m, \notag \\
			& & & \qquad z^+ \in \{0, 1\}^m, \; z^- \in \{0, 1\}^m \notag \\
			&\text{s.t.} & & f = p - B^+ A^+ \eta^+ - B^- A^- \eta^- 
			\label{prob2:f} \\ 
			& & & \eta_e^+ = \sin(y_e - \phi_{s(e), t(e)}), \; \forall e \in \mathcal E 
			\label{prob2:eta-plus} \\
			& & & \eta_e^- = -\sin(y_e + \phi_{t(e), s(e)}), \; \forall e \in \mathcal E 
			\label{prob2:eta-minus} \\
			& & & |y| \le \gamma \label{prob2:y-bound} \\ 
			& & & z^+_e = 1 \implies y_e = \gamma_e ~\text{and}~ \label{prob2:ind-plus} \\
			& & & \qquad d_{s(e)}^{-1} f_{s(e)} - d_{t(e)}^{-1} f_{t(e)} \ge 0, \; 
			\forall e \in \mathcal E \notag \\
			& & & z^-_e = 1 \implies y_e = -\gamma_e ~\text{and}~ \label{prob2:ind-minus} 
			\\
			& & & \qquad d_{s(e)}^{-1} f_{s(e)} - d_{t(e)}^{-1} f_{t(e)} \le 0, \; 
			\forall e \in \mathcal E \notag \\
			& & & \sum_{e \in \mathcal E} z_e^+ + z_e^- = 1 \label{prob2:sos1} 
		\end{align} 
		\end{subequations}
		If the problem is infeasible, we define $\widehat V(\gamma) = +\infty$.
		\label{prob:milp}
	\end{problem}

	\noindent 
	Recall that $s(e)$ and $t(e)$ represent the arbitrary ``source'' and ``target'' nodes 
	of each $e \in \mathcal E$. To express constraint \eqref{prob2:f} succinctly, 
	we decompose the incidence 
	matrix $B$ into two matrices $B^+, B^- \in \{0, 1\}^{n \times m}$, where $(B^+)_{i, 
	e} = 1$ if and only if $s(e) = i$, and $(B^-)_{i, e} = 1$ if and only if $t(e) = i$, 
	so that $B = B^+ - B^-$. We also define two diagonal matrices $A^+ = \diag\{ a_{s(e), 
	t(e)}, \; \forall e \in \mathcal E\}$ and $A^- = \diag\{ a_{t(e), s(e)}, \; \forall e 
	\in \mathcal E\}$. Constraints \eqref{prob2:ind-plus} and \eqref{prob2:ind-minus} are 
	indicator 
	constraints: if $z_e^+ = 1$, then 
	the constraints $y_e = \gamma_e$ and $d_{s(e)}^{-1} f_{s(e)} - d_{t(e)}^{-1} 
	f_{t(e)} \ge 0$ become ``active,'' but these constraints do not apply if $z_e^+ = 0$. 
	Indicator constraints are easily encoded in the MILP framework 
	\cite{PB-AL-AT-SW:15}, and many MILP solvers allow indicator constraints to 
	be supplied explicitly.\footnote{CPLEX 12.9 supports explicit 
	\href{https://www.ibm.com/support/knowledgecenter/SSSA5P_12.9.0/ilog.odms.cplex.help/CPLEX/UsrMan/topics/discr_optim/indicator_constr/01_indicators_title_synopsis.html}
		{indicator constraints}. Similarly, the Python interface to Gurobi 9 provides the 
		method 
		\href{https://www.gurobi.com/documentation/9.0/refman/py_model_agc_indicator.html}
	{\texttt{Model.addGenConstrIndicator()}}. Both links accessed 8/9/2020.}
 
 	Problem \ref{prob:milp} relaxes Problem \ref{prob:minmax} by optimizing the 
 	vector of counterclockwise differences $y = B^\top \theta$ directly, instead of 
 	optimizing $\theta \in \mathbb T^n$. Given this interpretation of $y$, constraints 
 	\eqref{prob2:f}--\eqref{prob2:eta-minus} ensure that $f = f(\theta)$ and that the 
 	cost function \eqref{prob2:cost} is equal to $V(\theta)$. Constraint 
	\eqref{prob2:y-bound} guarantees that $\theta \in \Delta(\gamma)$, and 
	\eqref{prob2:ind-plus}--\eqref{prob2:sos1} ensure that the underlying $\theta$ is on 
	the ``outward-pointing'' boundary of $S$. 
	
	The most important property of Problem 
	\ref{prob:milp} is that it yields a lower 
	bound to $V^*(\partial \Delta(\gamma))$:
	\begin{lemma}[Problem \ref{prob:milp} is a Relaxation]
		Let $\gamma \in (0, \gamma^*]$, and let $S = \Delta(\gamma)$. 
		The solutions to Problems \ref{prob:minmax} and \ref{prob:milp} are related by 
		$\widehat V(\gamma) \le V^*(\partial S)$, where equality holds if the underlying 
		graph $G$ is a tree.
		\label{lem:probs1}
	\end{lemma} 
	\noindent
	The proof is contained in Appendix \ref{appendix:prob}. Due to this bound, we can 
	replace the $\delta_0 < V(\Delta(\gamma))$ condition in Theorem~\ref{thm:general} 
	with the stricter (but computable) condition $\delta_0 < \widehat V(\gamma)$.
	
	\begin{theorem}[Computational Certificate] \label{thm:milp}
		Consider a trajectory $\theta(t)$ of \eqref{eq:dynamics} on any connected graph 
		$G$. Let $\gamma_0 = |B^\top 
		\theta(0)|$ and $\delta_0 = V(\theta(0))$ denote the initial angle 
		differences and initial max frequency deviation. If there exists a vector $\gamma 
		\in [\gamma_0, \gamma^*]$ such that 
		$\delta_0 < \widehat V(\gamma)$, then statements \ref{p:fi}--\ref{p:energy}
		from Theorem~\ref{thm:general} hold, with respect to the set $S = \Delta(\gamma)$.
	\end{theorem}
	
	\begin{proof}
		Let $S = \Delta(\gamma)$. By Lemma \ref{lem:probs1}, $\delta_0 < 
		\widehat V(\gamma) \le V^*(\partial S)$, so $\gamma$ and $S$ 
		satisfy the hypothesis of Theorem~\ref{thm:general}.
	\end{proof}

	\noindent
	Given a particular $\gamma \in [\gamma_0, \gamma^*]$, Theorem~\ref{thm:milp} provides 
	a certificate for transient stability and the other operating 
	constraints in Definition \ref{def:dp}, using only two properties of the initial 
	condition: the initial angle differences $\gamma_0$, and the initial maximal 
	frequency deviation $\delta_0$. Furthermore, Theorem~\ref{thm:milp} replaces Problem 
	\ref{prob:minmax} with Problem \ref{prob:milp}, which is more readily solved by 
	numerical methods. 
	
	But two issues still remain. The first problem with Theorem 
	\ref{thm:milp} is that it is conservative, since 
	Problem \ref{prob:milp} is a lower bound on Problem 
	\ref{prob:minmax}. This bound is only tight in acyclic 
	networks, and the gap between these two problems tends to increase with the number of 
	edges in the graph. In other words, denser graphs lead to more conservative 
	certificates provided by Theorem~\ref{thm:milp} (compared to Theorem 
	\ref{thm:general} applied to $S = 
	\Delta(\gamma)$). Closing this gap 
	with additional constraints requires some deeper analysis of the $n$-torus geometry, 
	which we postpone to Section \ref{sect:winding}. 
	
	The second issue is that 
	Problem \ref{prob:milp} is difficult to solve. It is 
	possible to tackle the problem using nonlinear programming techniques, but it is 
	faster and safer to relax the problem to obtain a lower bound, which we examine next.
	
	\subsection{MILP Certificate} 
	
	Problem \ref{prob:milp} is challenging to solve numerically, since it 
	contains the nonlinear equality constraints \eqref{prob2:eta-plus} and 
	\eqref{prob2:eta-minus}. 
	Furthermore, we must be careful about 
	using nonlinear solvers to estimate $\widehat V$. If the solver obtains a sub-optimal 
	solution, then this solution may exceed $V^*(\partial \Delta(\gamma))$, thereby 
	invalidating Theorem~\ref{thm:milp}. But \textit{any lower bound 
	on $\widehat V(\gamma)$ can be used in place of $\widehat V(\gamma)$ in Theorem 
	\ref{thm:milp}}. We can get a lower bound---while simultaneously making the 
	problem much easier to solve---by further relaxing Problem \ref{prob:milp} into a 
	MILP.
	
	\begin{figure}
		\begin{minipage}{0.48\linewidth} 
			\includegraphics[clip,width=\linewidth,trim={0.5in 0 0 
			0}]{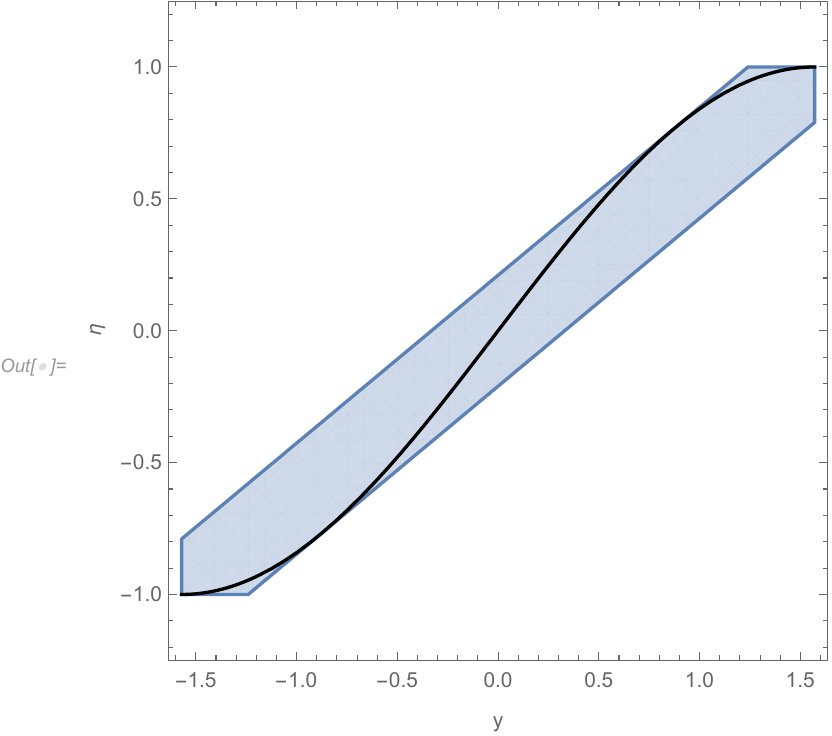}
		\end{minipage}
		\begin{minipage}{0.48\linewidth} 
			\includegraphics[clip,width=\linewidth,trim={0.5in 0 0 
				0}]{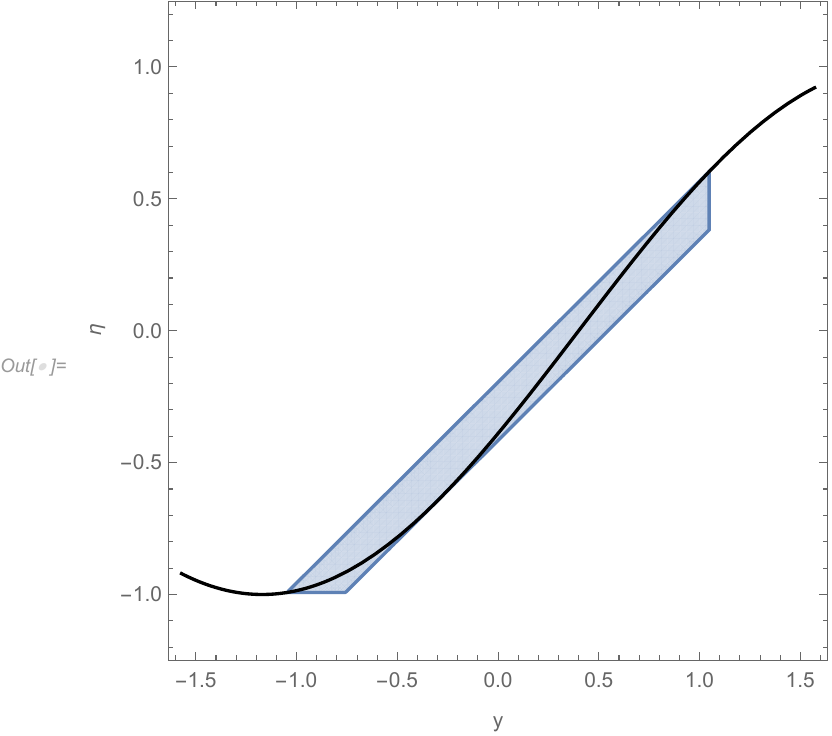}
		\end{minipage}
		\caption{Sample relaxations to the constraint $\eta^+_e = \sin(y_e - 
		\phi_{s(e),t(e)})$ on $y_e \in [-\gamma_e, \gamma_e]$ using bounding polytopes. 
		In the left example, $\phi_{s(e), t(e)} = 0$ and $\gamma_e = \frac \pi 2$. In the 
		right example, $\phi_{s(e), t(e)} = 0.4$ and $\gamma_e = \frac \pi 3$.} 
		\label{fig:relaxation}
	\end{figure}

	The relaxation is conceptually simple; all we need to do is replace 
	\eqref{prob2:eta-plus} and 
	\eqref{prob2:eta-minus} with linear and integer constraints. The general idea is to 
	find a polytope or a union of polytopes that contain the sine curve. Then these 
	bounds can be encoded within the MILP and substituted in for 
	\eqref{prob2:eta-plus} and \eqref{prob2:eta-minus}. The process of constructing these 
	polytopes is a messy exercise in elementary geometry, so we do not go into details 
	here. Instead, we present two examples in Figure \ref{fig:relaxation}, both of which 
	relax the sine constraints with four linear bounds (although tighter and more 
	complicated bounds are clearly possible.) In fact, one can achieve arbitrary 
	precision in the relaxed sine constraints by using piecewise-linear bounds, at the 
	expense of additional binary variables and slower computation.

	Replacing \eqref{prob2:eta-plus} and \eqref{prob2:eta-minus} with the polytope 
	relaxation turns Problem \ref{prob:milp} into a MILP. This 
	MILP can be solved with standard software like Gurobi, CPLEX, or MATLAB. The 
	solution is a lower bound on $\widehat V(\gamma)$, which can safely be used in 
	place of $\widehat V(\gamma)$ in Theorem~\ref{thm:milp}. We also note that this MILP 
	is computationally tractable, since the binary variables essentially split the 
	problem into $2m$ linear programming sub-problems, each corresponding to one of the 
	$2m$ faces of $\Delta(\gamma)$. Informally, Problem 
	\ref{prob:milp} is no more complex than a collection of $2m$ linear programs.
}
	 
 	\section{Improved Guarantees for Meshed Networks}
	\label{sect:winding} 
\rev{
	In the previous section, we found transient guarantees based on two properties of the 
	initial condition, $\gamma_0$ and $\delta_0$. But if $G$ is a cyclic topology (i.e., 
	the graph is not a tree), we can make use of an additional property of the initial 
	condition: its \textit{winding vector}, $u_0$. 
	Winding vectors have recently gained attention in the study of power transmission 
	networks due to their relationship with loop flows \cite{NJ-AK:03, TC-RD-IA-PJ:16}, 
	i.e., net flows of power around cycles in the network. In \cite{SJ-EYH-KDS-FB:18jv3}, 
	we partitioned the $n$-torus into equivalence classes of winding vectors, and we 
	showed that a wide class of network systems with phase-valued states (including the 
	Kuramoto model) have at most one equilibrium point within each equivalence class. 
	Since winding vectors contain enough information to uniquely characterize 
	equilibrium points of the system, it seems they may also provide 
	information about the transient behavior. 
	
	In this section, we briefly review concepts related to the winding partition. We 
	then apply these concepts to the transient stability problem at hand, using 
	knowledge of the initial winding vector $u_0$ to obtain less-conservative 
	certificates from Theorem~\ref{thm:general} and completely close the gap between 
	Problems 
	\ref{prob:minmax} and \ref{prob:milp}. For simplicity, we assume throughout this 
	section that $G$ contains at least one cycle. (Otherwise $G$ is a tree, in which 
	case there is no gap between Problems \ref{prob:minmax} and \ref{prob:milp} to begin 
	with.)
		
	\subsection{Winding Partition of the $n$-Torus} \label{sect:winding-intro}
	
	\paragraph*{Preliminaries} We start with some preliminaries on algebraic graph theory 
	and its application to graph cycles. We refer the reader to \cite[\S 9.3]{FB:20} for 
	a more detailed discussion of these concepts. A \textit{simple cycle} in $G$ is a 
	sequence of consecutive nodes, where the first and last nodes are identical, but all 
	other nodes are distinct. Given a simple cycle $\sigma = (i_1, i_2, \dots, 
	i_{n_\sigma}, i_1)$, the \textit{cycle vector} $v_\sigma \in \real^m$ is defined with 
	respect to the incidence matrix $B$ by
	\begin{align*}
	{(v_\sigma)}_e =
	\begin{cases}
	+1,\quad & \mbox{if the edge $e$ is traversed positively by $\sigma$,}
	\\
	-1, \quad & \mbox{if the edge $e$ is traversed negatively by $\sigma$,} \\
	0, & \mbox{otherwise}.
	\end{cases}
	\end{align*}
	for each $e \in \mathcal E$. More formally, given an adjacent pair of nodes $i_j, 
	i_{j + 1}$ in $\sigma$, we say that $\sigma$ traverses the edge $\{i_j, i_{j + 1}\}$ 
	\textit{positively} if $B_{i_j, e} = +1$ and $B_{i_{j+1}, e} = -1$; otherwise, it 
	traverses the edge \textit{negatively}. The set of cycle vectors for all simple 
	cycles in $G$ span a vector space, called the \textit{cycle space} of $G$. A set of 
	simple cycles $\Sigma$ is called a \textit{cycle basis} if the cycle vectors 
	corresponding to elements of $\Sigma$ are a basis for the cycle space. A cycle basis 
	$\Sigma = \{\sigma_1, \sigma_2, \dots, \sigma_{|\Sigma|} \}$ can be 
	encoded in a cycle-edge incidence matrix $C_\Sigma \in \{-1, 0, 1\}^{|\Sigma| \times 
	m}$, where
	\[
		C_\Sigma = \begin{pmatrix} v_{\sigma_1} & v_{\sigma_2} & \cdots &		
		v_{\sigma_{|\Sigma|}} \end{pmatrix}^\top
	\] 
	Clearly $\Img(C_\Sigma^\top)$ is the cycle space. Furthermore, because $G$ is a 
	connected graph, the dimension of the cycle space is $|\Sigma| = m - n + 1$.  
	
	\paragraph*{Winding Vectors and Winding Partition} We now review basic definitions 
	regarding winding vectors and the winding partition. The partition 
	divides the $n$-torus into equivalence classes induced by an underlying graph $G$. 
	These equivalence classes are defined by how many times the phase differences across 
	basis cycles of $G$ ``wind'' around the unit circle:
}
	
	\begin{definition}[Winding Numbers, Vectors, and Cells] \label{def:winding}
		Let $\theta \in \mathbb T^n$. Given any simple cycle $\sigma$ in $G$ with 
		$n_\sigma$ nodes, the \textit{winding number} of $\theta$ along $\sigma$ is
		\begin{equation} \label{eq:winding-num}
		w_\sigma(\theta) = \frac{1}{2\pi} \sum_{i = 1}^{n_\sigma} d_{\rm cc}(\theta_i, 
		\theta_{i + 1})
		\end{equation}
		where the nodes in $\sigma$ are indexed $\sigma = (1, \dots, n_{\sigma}, 1)$ and 
		$\theta_{n_\sigma + 1} = \theta_1$. Given a cycle basis $\Sigma$ of $G$, the 
		winding vector of $\theta$ along $\Sigma$ is the vector
		\begin{equation} \label{eq:winding-vec}
		w_\Sigma(\theta) = \begin{pmatrix} w_{\sigma_1}(\theta) & w_{\sigma_2}(\theta) & 
		\cdots & w_{\sigma_{|\Sigma|}}(\theta) \end{pmatrix}^\top
		\end{equation}
		For every winding vector $u \in w_\Sigma(\mathbb T^n)$, the $u$-winding cell is 
		the equivalence class
		\begin{equation} \label{eq:winding-cell}
		\Omega_u = \{ \theta \in \mathbb T^n : w_\Sigma(\theta) =u\}
		\end{equation} 
		\label{def:wc}
	\end{definition}
\rev{
	\noindent
	We note that winding cells were introduced in \cite{SJ-EYH-KDS-FB:18jv3}, but winding 
	vectors have been used in the study of power flows since \cite{NJ-AK:03}. The reader 
	has likely encountered a similar concept of ``winding number'' from
	interpreting Nyquist plots.
}
	
	\begin{figure}
		\includegraphics[width=\linewidth]{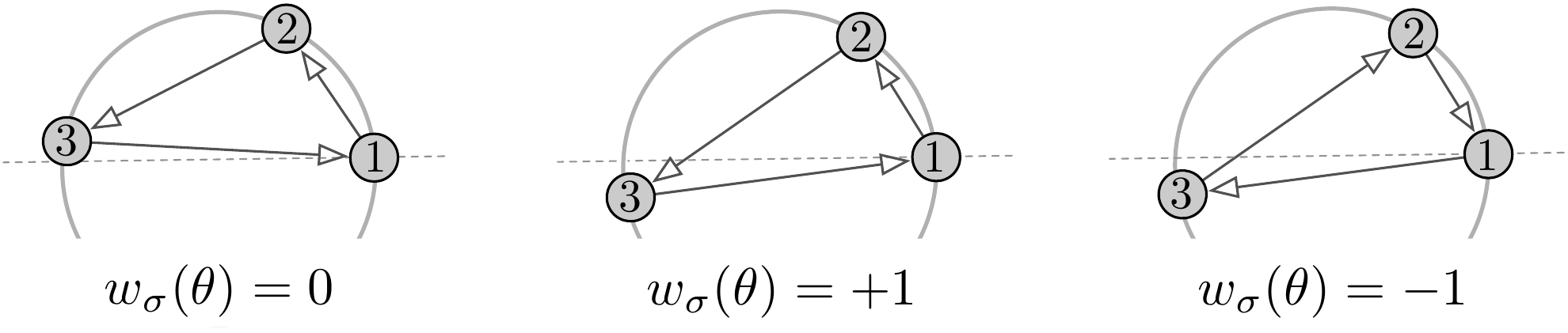}
		\includegraphics[width=\linewidth]{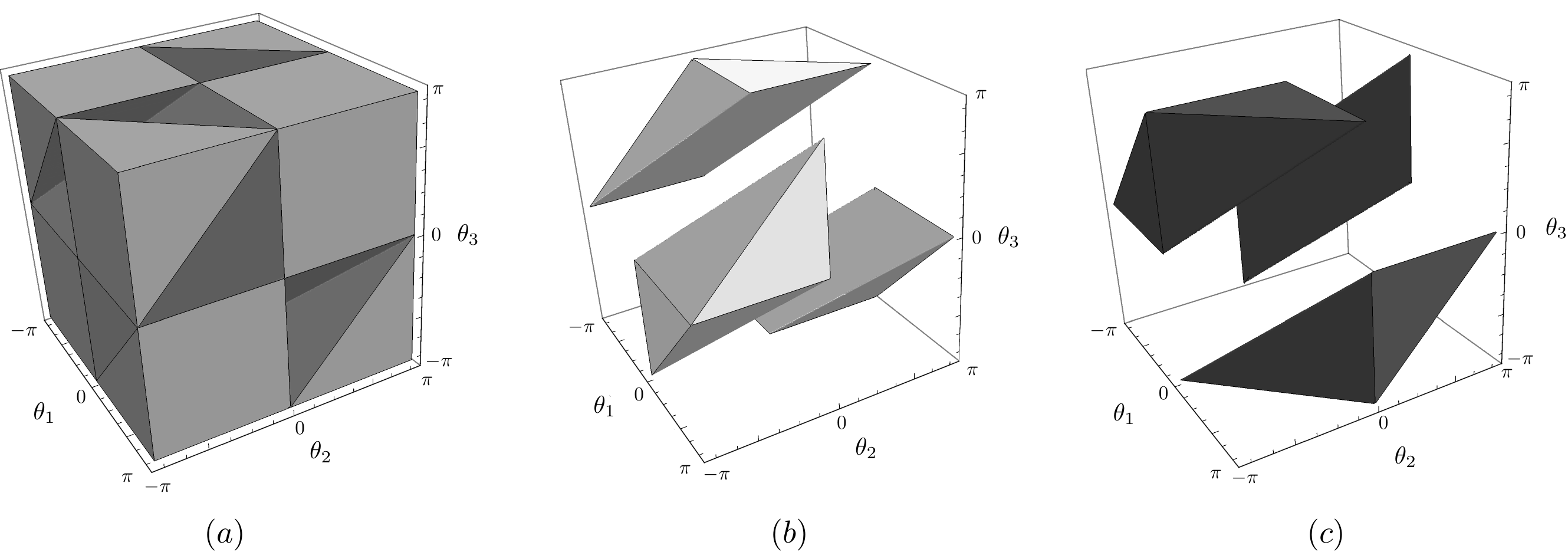}
		\caption{Possible winding numbers for any 3-torus state on the 3-cycle (top), and 
			the winding cells in $\mathbb T^3$ corresponding to each of these possible 
			winding numbers (bottom).}
		\label{fig:3cycle_windingnum}
	\end{figure}
	
	Winding vectors are always integer-valued, a property which is analogous to 
	Kirchoff's voltage law (KVL). For real-valued nodal potentials, KVL guarantees 
	that potential differences sum to zero around any cycle. Similarly, phase 
	differences (in the sense of counter-clockwise arc length) sum to an integer 
	multiple of $2\pi$ around any cycle. For example, suppose that $G$ is the 
	triangle graph, consisting of a 
	single cycle $\sigma = (1, 2, 3, 1)$. Let $\theta \in \mathbb T^3$. If there is an 
	arc of length $\pi$ that contains $\theta_1, \theta_2, \theta_3$, then 
	$w_\sigma(\theta) = 0$. Otherwise, $w_\sigma(\theta) = \pm 1$. Figure 
	\ref{fig:3cycle_windingnum} (top) illustrates these three possible winding numbers, 
	based on the configuration of phases around the cycle. Meanwhile, Figure 
	\ref{fig:3cycle_windingnum} (bottom) illustrates $\Omega_u$ for each $u = -1, 0, 1$.

\rev{
	The cycle basis of a graph is often non-unique, and each valid cycle basis $\Sigma$ 
	leads to a different definition of the winding vector $w_\Sigma(\theta)$. However, it 
	turns out that the equivalence classes of $w_\Sigma$, namely the winding cells, do 
	\textit{not} depend on the particular choice of cycle basis. Given 
	two cycle bases $\Sigma$ and $\Sigma'$, every winding cell based on $w_\Sigma$ is 
	identical to a winding cell based on $w_{\Sigma'}$.
		
	\paragraph*{Physical Interpretation of Winding Vectors}
	
	Winding vectors are closely connected to net flows of active power around cycles in 
	the network. These flows, called loop flows, are of considerable interest to the 
	power systems community because they do not deliver useful power and can jeopardize 
	system stability. For example, flows around the Lake Erie Loop were a major factor in 
	the 2003 Northeast Blackout \cite{NERC-04}. Rigorous connections between loop 
	flows and winding vectors are discussed in detail in \cite{NJ-AK:03, TC-RD-IA-PJ:16, 
	SJ-EYH-KDS-FB:18jv3}.  We here provide some basic intuition  and show how to 
	measure the initial winding vector using line flows, instead of the full state 
	$\theta(0)$.
	
	Consider the active power flows across each line given by \eqref{p:line}. We assume 
	that these flows are measurable, so that $p^{\rm line}$ is known, even if the state 
	$\theta$ is not. If $\theta \in \Delta(\gamma^*)$, then \eqref{eq:winding-num} can be 
	written
	\begin{equation} \label{eq:w-measure}
		w_\sigma(\theta) = \frac{1}{2\pi} \sum_{i = 1}^{n_\sigma} \arcsin\left( 
		\frac{p^{\rm line}_{i, i + 1} - \tilde a_{i, i+1}}{a_{i,i+1}}\right) + 
		\phi_{i,i+1}
	\end{equation}
	If we ignore shunt and series losses by setting $\tilde a_{i,i+1} = 0$ and 
	$\phi_{i,i+1} = 0$, and we expand the arcsine function about the origin, we obtain 
	\[
		w_\sigma(\theta) = \frac{1}{2\pi} \sum_{i=1}^{n_\sigma} \frac{p_{i,i+1}^{\rm 
		line}}{a_{i,i+1}} + O((p_{i,i+1}^{\rm line})^3)
	\] 
	The quantity $a_{i,i+1}^{-1} p_{i,i+1}^{\rm line}$ is a normalized line flow, scaled 
	by the capacity of the line. Thus, up to second order, the winding number is a 
	normalized loop flow (at least in the case of short, lossless transmission lines). 
	While somewhat informal, this analysis suggests that winding vectors are a quantized 
	measure of these loop flows. We can also use \eqref{eq:w-measure} to infer the 
	winding vector from line flow data. 
	Therefore, like the vector of angle differences $\gamma_0$, we can identify the 
	initial winding vector $u_0$ using measurements of line flows instead of the full 
	state $\theta(0)$. 	
 
	\subsection{Improved Certificates}
	
	We have previously seen (in Theorems \ref{thm:general} and \ref{thm:milp}) how 
	to certify transient stability and other desirable properties from the initial 
	condition, using the measurable quantities $\gamma_0$ and $\delta_0$ instead of 
	the full state $\theta(0)$. As with $\gamma_0$ and $\delta_0$, the initial winding 
	vector $u_0 = w_\Sigma(\theta(0))$ provides additional information that we can 
	exploit to make guarantees about the transient. In the remainder of this section, we 
	will show how to use the winding vector to obtain better certificates out of Theorems 
	\ref{thm:general} and \ref{thm:milp}, replacing the antecedents of these theorems 
	with less-conservative conditions.
	
	The new conditions are straightforward to 
	state and prove. We modify Theorem~\ref{thm:general} to search over sets of the form 
	$\Delta(\gamma_0) \cap \Omega_{u_0} \subseteq S \subseteq \Delta(\gamma) \cap 
	\Omega_{u_0}$ instead of $\Delta(\gamma_0) \subseteq S \subseteq \Delta(\gamma)$. 
	Reducing the lower bound from $\Delta(\gamma_0)$ to $\Delta(\gamma_0) \cap 
	\Omega_{u_0}$ directly incorporates $u_0$ and results in a larger search space 
	for $S$, thereby expanding the set of cases that satisfy the antecedent of Theorem 
	\ref{thm:general}. Shrinking the upper bound from $\Delta(\gamma)$ to $\Delta(\gamma) 
	\cap \Omega_{u_0}$ is not strictly necessary, but as we will see later on, we can 
	always find an optimal $S$ within this smaller upper bound. Similarly, we will modify 
	Theorem~\ref{thm:milp} by adding a constraint to Problem \ref{prob:milp} 
	that forces the optimum to reside within $\Omega_{u_0}$:

	\begin{problem}[Min-Max Frequency Deviation, Exact] \label{prob:exact}
		Let $\gamma \in (0, \gamma^*]$ and $u \in \Img(w_\Sigma)$. We define $\widehat 
		V(\gamma, u)$ as the minimum value of Problem \ref{prob:milp}, under the 
		additional constraint $C_\Sigma y = 2\pi u$. If the problem is infeasible, we 
		define $\widehat V(\gamma, u) = +\infty$. 
	\end{problem} 

	\noindent
	The additional $C_\Sigma y = 2\pi u$ constraint confines the solution to $\Omega_u$, 
	completely closing the gap between Problems \ref{prob:minmax} and \ref{prob:milp}:

	\begin{lemma}[Relations of Minima] \label{lem:probs2}
		Let $\gamma \in (0, \gamma^*]$, let $u \in \Img(w_\Sigma)$, and let $S = 
		\Delta(\gamma) \cap \Omega_u$. The solutions to Problems \ref{prob:minmax}, 
		\ref{prob:milp}, and \ref{prob:exact} are related by 
		\[
			\widehat V(\gamma) \le \widehat V(\gamma, u) = 
			V^*(\partial S).
		\] 
	\end{lemma}

	\noindent
	The proof is contained in Appendix \ref{appendix:prob}. We can now state the new 
	transient stability certificates that account for the initial winding vector. 
	
	\begin{theorem}[Certificates with Winding Vectors] \label{thm:winding}
		Consider a trajectory $\theta(t)$ of \eqref{eq:dynamics}, and let $\Sigma$ 
		be a cycle basis of the underlying graph. Let $\gamma_0 = |B^\top \theta(0)|$, 
		$\delta_0 = V(\theta(0))$, and $u_0 = w_\Sigma(\theta(0))$ denote the 
		initial angle differences, max frequency deviation, and winding 
		vector. Consider the following two conditions:
		\begin{enumerate}[label=(\alph*)]
			\item \label{cond:set}
			There exist a vector $\gamma \in [\gamma_0, \gamma^*]$ and a set 
			$\Delta(\gamma_0) \cap \Omega_{u_0} \subseteq S 
			\subseteq \Delta(\gamma) \cap \Omega_{u_0}$ such 
			that $\delta_0 < V^*(\partial S)$. 
			\item \label{cond:lp}
			There exists a vector $\gamma \in [\gamma_0, \gamma^*]$
			such that $\delta_0 < \widehat V(\gamma, u_0)$.
		\end{enumerate}
		If either \ref{cond:set} or \ref{cond:lp} are true, then statements 
		\ref{p:fi}--\ref{p:energy} from Theorem~\ref{thm:general} hold, with respect to 
		$S$ from condition \ref{cond:set} or $S = \Delta(\gamma) \cap \Omega_{u_0}$ from 
		condition \ref{cond:lp}.
	\end{theorem} 

	\begin{proof}
		The proof that statements \ref{p:fi}--\ref{p:energy} follow from \ref{cond:set} 
		is identical to the proof of Theorem~\ref{thm:general}, since the new upper and 
		lower bounds on $S$ do not impact the argument for statement \ref{p:fi}, and 
		(because $S$ is still contained within 
		$\Delta(\gamma)$) they have no bearing on statements 
		\ref{p:stability}--\ref{p:energy}. We can use this result from condition 
		\ref{cond:set} to prove that \ref{p:fi}--\ref{p:energy} follow from condition
		\ref{cond:lp}. Let $S = \Delta(\gamma) \cap \Omega_{u_0}$, and observe that 
		$\delta_0 < \widehat V(\gamma, u_0) \le V^*(\partial S)$ due to 
		Lemma \ref{lem:probs2}. Then $S$ satisfies \ref{cond:set}, so all of the 
		statements hold.
	\end{proof} 
 
	It is straightforward to show that these new conditions which incorporate $u_0$ are 
	valid certificates for transient stability, but this is not enough---if we are to go 
	to the trouble of measuring $u_0$, we would like the assurance that this additional 
	information actually leads to better transient stability certificates. With some 
	simple but careful reasoning about the winding partition, we can see that 
	\ref{cond:set} and \ref{cond:lp} are less-conservative versions of the antecedents to 
	Theorems \ref{thm:general} and \ref{thm:milp}, respectively:  
		
	\begin{theorem}[Theorem~\ref{thm:winding} is less conservative than Theorems 
	\ref{thm:general} and \ref{thm:milp}]
	\label{thm:conservative}
		Consider a trajectory $\theta(t)$ of \eqref{eq:dynamics}, let $\Sigma$ 
		be a cycle basis of the underlying graph, and let $\gamma_0 = |B^\top 
		\theta(0)|$, $\delta_0 = V(\theta(0))$, and $u_0 = w_\Sigma(\theta(0))$. The 
		following are true:
		\begin{enumerate}
			\item \label{item:3-bound}
			If the hypothesis of Theorem~\ref{thm:general} is satisfied, i.e., if there 
			exist a vector $\gamma \in [\gamma_0, \gamma^*]$ and a set $\Delta(\gamma_0) 
			\subseteq S \subseteq \Delta(\gamma)$ such that $\delta_0 < V^*(\partial S)$, 
			then the set $S' = S \cap \Omega_{u_0}$ satisfies condition \ref{cond:set} of 
			Theorem~\ref{thm:winding}.
			\item \label{item:5-bound}
			If the hypothesis of Theorem~\ref{thm:milp} is satisfied, i.e., if 
			there exists a vector $\gamma \in [\gamma_0, \gamma^*]$ such that $\delta_0 < 
			\widehat V(\gamma)$, then $\gamma$ satisfies condition \ref{cond:lp} of 
			Theorem~\ref{thm:winding}.
		\end{enumerate}
	\end{theorem} 

	\begin{proof}	
		To prove \ref{item:3-bound}, it is sufficient to show that $V^*(\partial 
		S') \ge V^*(\partial S)$, for which it is sufficient to show that 
		$\partial (S \cap \Omega_{u_0}) \subseteq \partial S$. Because the winding cells 
		partition $\mathbb T^n$, each of the sets $\Delta(\gamma) \cap \Omega_u$ are 
		disjoint. In fact, because $\gamma < \pi \vect 1_m$, the boundaries of these sets 
		are non-overlapping. Since $S \subseteq \Delta(\gamma)$, we may
		conclude that $\partial S$ itself is partitioned into non-overlapping pieces 
		$\partial(S \cap \Omega_{u_0})$; hence $\partial (S \cap \Omega_{u_0}) \subseteq 
		\partial S$. Similarly, for \ref{item:5-bound} it is sufficient to show that 
		$\widehat V(\gamma, u_0) \ge \widehat V(\gamma)$, which we have 
		from Lemma \ref{lem:probs2}.
	\end{proof}

	The initial winding vector provides an additional bit of information about the 
	initial state $\theta(0)$, and like the vector of initial angle differences 
	$\gamma_0$, the initial winding vector $u_0$ can be inferred from measurements of 
	active power flows. With knowledge of $u_0$, we can replace the set-theoretic 
	certificate in Theorem~\ref{thm:general} and the MILP certificate in 
	Theorem~\ref{thm:milp} with the less-conservative conditions \ref{cond:set} and 
	\ref{cond:lp} of Theorem~\ref{thm:winding}. 
}
 
\rev{
	\section{Quantifying Robustness}
	\label{sect:robust}
	
	An important task in power systems control is understanding how robust an operating 
	point is to disturbances. It is straightforward to study the effects of a 
	\textit{particular} disturbance using simulation, but simulating a comprehensive set 
	of contingencies (or combinations thereof) is time consuming. Our transient 
	stability certificates can aid with this analysis by quantifying the scale of 
	disturbances to which an operating point is robust. 
	
	In this section, we consider a DCMG that is operating at a synchronous state 
	$\theta_0$. At time $t = 0$, certain model parameters undergo an instantaneous 
	perturbation---nominal injections change, for example, or a drop in nodal voltages or 
	branch admittances occurs. The initial condition $\theta_0$ is no longer a 
	synchronous state in the ``post-fault'' model. If the system is sufficiently 
	resilient, then the post-fault transient will settle back down to a synchronous 
	state, and none of the engineering constraints will be violated in the process---but 
	this is not always the case. We will construct a sufficient condition for post-fault 
	transient stability, based on the scale of the perturbations to model parameters.
	 
	\paragraph*{Numerical Case Study}
	Throughout the section, we will illustrate our results using numerical examples from 
	the IEEE-RTS 24-bus test case 
	\cite{CG-PW-PA-RA-MB-RB-QC-CF-SH-SK-WL-RM-DP-NR-DR-AS-MS-CS:99}. We 
	parameterized \eqref{eq:dynamics} using branch and bus values from this test case, 
	and we selected the 
	initial voltage angles $\theta_0 \in \mathbb T^n$, voltage magnitudes, and nominal 
	power injections by solving for the optimal power flow in MATPOWER. For simplicity, 
	we chose uniform droop coefficients of 
	$10~\text{pu}\cdot\text{s}$ and a uniform nominal frequency of $60~s^{-1}$. We will 
	refer to this model as the ``pre-fault'' model. Note that $\theta_0$ is a synchronous 
	equilibrium of the pre-fault model (since it solves the active power flow 
	equations with nominal injections), and the initial winding vector is $u_0 = \vect 
	0_{11}$ (corresponding to the winding cell with minimal loop flows).
	
	\paragraph*{Code}
	The code that we used to generate numerical results in this section is publicly 
	available at \url{https://github.com/KevinDalySmith/DCMG-transient-stability}. The 
	optimization problems are implemented using the Python interface to Gurobi 9.0, so a 
	local installation of Gurobi and an active license are needed to run it.
		
	\subsection{Evaluating Post-Fault Transient Stability}
	
	\begin{figure} 
		\centering
		\includegraphics[width=0.9\linewidth, clip, trim={1.5cm 0 0 0}]{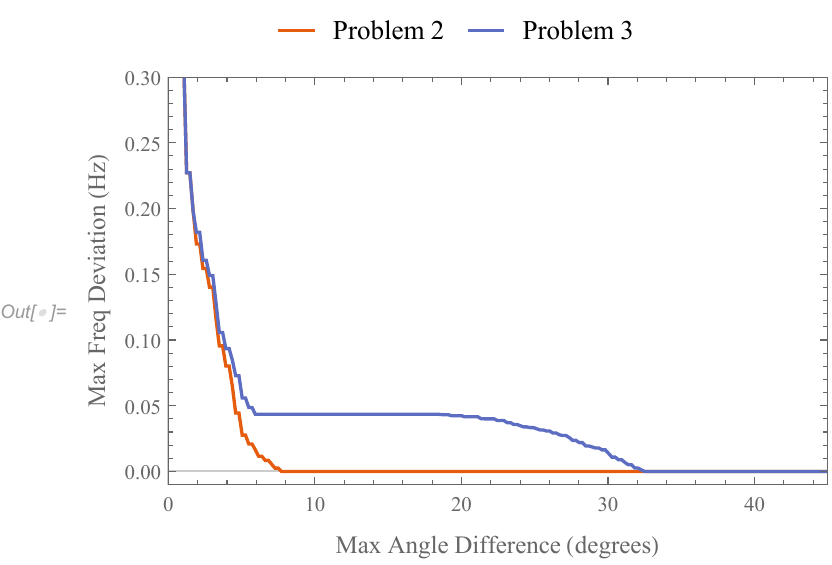} 
		\caption{Transient stability certification in the IEEE 24-bus test case. The 
			horizontal axis is the max angle difference $||B^\top \theta_0||_\infty$ of 
			the initial condition, and the vertical axis is the largest max frequency 
			deviation $V(\theta_0)$ for which transient stability is certified. The lower 
			curve is computed using $\widehat V(\gamma)$, and the upper curve is computed 
			using $\widehat V(\gamma, \vect 0_{11})$.}
		\label{fig:region}
	\end{figure} 
	
	We begin by examining how Theorems \ref{thm:milp} and \ref{thm:winding} apply to 
	the problem of quantifying system robustness. Both of these theorems certify 
	transient stability if the post-fault initial condition (i.e., the pre-fault 
	synchronous state) is sufficiently close to a post-fault synchronous state, as 
	measured by the initial max frequency deviation, $\delta_0 = V(\theta_0)$. Then 
	transient stability is certified if there exists any $\gamma \in [\gamma_0, 
	\gamma^*]$ such that $\delta_0 < \widehat V(\gamma)$ or $\delta_0 < \widehat 
	V(\gamma, u_0)$, as in the following example.
	
	\begin{example}[Certificates in the 24-Bus System] \label{ex:24}
		To illustrate Theorems \ref{thm:milp} and \ref{thm:winding} in the 24-bus system, 
		we randomly select a set of 100 test points $\Gamma \subset [\vect 0_m, \; 
		\gamma^*]$ and evaluate $\widehat V(\gamma)$ and $\widehat V(\gamma, u_0)$ at 
		each $\gamma \in \Gamma$, with $u_0 = \vect 0_{11}$. Consider an arbitrary 
		initial condition with a maximum angle difference $\bar \gamma = ||B^\top 
		\theta(0)||_\infty$ and initial frequency deviation $\delta_0 = V(\theta(0))$.	
		Theorem~\ref{thm:milp} certifies transient stability of the resulting trajectory 
		if 
		\[
			\delta_0 < \max_{\gamma \in \Gamma} \left\{ \widehat V(\gamma) : \gamma \ge 
			\bar \gamma \vect 1_m \right\}.
		\]
		Under the additional assumption that $w_\Sigma(\theta(0)) = \vect 
		0_{11}$, Theorem~\ref{thm:winding} certifies transient stability if
		\[
			\delta_0 < \max_{\gamma \in \Gamma} \left\{ \widehat V(\gamma, \vect 0_{11}) 
			: \gamma \ge \bar \gamma \vect 1_m \right\}.
		\]
		Figure \ref{fig:region} plots both of these conditions. Each curve plots the 
		right-hand side of the preceding inequalities as a function of $\bar \gamma$ 
		(using polytopic relaxations to the sine constraints). 	
		For any initial condition corresponding to a point $(\bar \gamma, \delta_0)$ 
		below the curve, transient stability is certified. 
		
		The curve maximizing $\widehat V(\gamma)$ is significantly lower 
		than the curve maximizing $\widehat V(\gamma, \vect 0_{11})$, i.e., the 
		transient stability condition from Theorem~\ref{thm:milp} is more 
		conservative than that from Theorem~\ref{thm:winding} (as guaranteed by Theorem 
		\ref{thm:conservative}). This plot makes a strong case for incorporating 
		information from the initial winding vector---without it, only very small angle 
		disturbances are certified in the 24-bus system. 
	\end{example}

	\noindent
	In order to certify transient stability after a fault, we must ensure that the 
	initial post-fault frequency deviation is below the \text critical threshold. One 
	approach is to follow the procedure of Example \ref{ex:24}: generate a plot similar 
	to Figure \ref{fig:region} using the post-fault parameters, and check whether or not 
	$\theta_0$ corresponds to a point below the curve. But this approach is cumbersome 
	when considering a large number of contingencies, and it offers little advantage over 
	simulation.
	
	A much more efficient approach, similar to that in \cite{TLV-KT:17}, is to 
	define the ``size'' of a general disturbance and establish a threshold below 
	which transient stability is certified in all disturbances that are ``smaller'' than 
	the threshold. A natural way 
	to define the size of a disturbance is to quantify its effect on the frequency 
	deviation vector from \eqref{def:freq-deviation}. Suppose that $v: \mathbb T^n \to 
	\mathbb R^n$ is the frequency deviation vector field defined with the pre-fault model 
	parameters, and similarly, let $\bar v$ be the vector field defined with the 
	post-fault parameters. If we can bound the difference $\xi(\theta) = \bar v(\theta) - 
	v(\theta)$, then we can bound the solutions to Problems \ref{prob:milp} and 
	\ref{prob:exact} after the disturbance based on their solutions before the 
	disturbance:

	\begin{theorem}[Robustness to Parameter Changes] \label{thm:robust}
		Consider the model \eqref{eq:dynamics}, and let $v: \mathbb T^n \to \mathbb R^n$ 
		be the associated frequency deviation vector. Let $\theta_0 \in \mathbb T^n$ be a 
		state for which $v(\theta_0) = 0$, i.e., for which all nodal frequencies are 
		identical to $\omega^*$. After some perturbation in model parameters, suppose 
		that the new frequency deviation vector is given by $\bar v(\theta) = v(\theta) + 
		\xi(\theta)$, and let $\gamma_0 = 
		|B^\top \theta_0|$ and $u_0 = w_\Sigma(\theta_0)$ in the post-fault model. If 
		there exists $\gamma \in [\gamma_0, \gamma^*]$ such that 
		\begin{equation} \label{eq:perturb} 
			||\xi(\theta_0)||_\infty + \max_{\theta \in \partial S} 
			||\xi(\theta)||_\infty < \min_{\theta \in \partial S} ||v(\theta)||_\infty
		\end{equation} 
		where either $S = \Delta(\gamma)$ or $S = \Delta(\gamma) \cap \Omega_{u_0}$, then 
		the trajectory of the 
		perturbed model starting from $\theta_0$ satisfies statements 
		\ref{p:fi}--\ref{p:energy} from Theorem~\ref{thm:general} with respect to $S$.
	\end{theorem}

	\begin{proof}
		Let $\Theta \subseteq \partial S$ be the feasible set of Problem  
		\ref{prob:minmax} 
		evaluated on the perturbed model, i.e., the set of points $\theta \in \partial S$ 
		such that $D^{-1} f(\theta)$ is pointed outward from $S$. Then the solution to 
		Problem \ref{prob:minmax} (evaluated on the perturbed model) is 
		\begin{align*}
			V^*(\partial S) &= \min_{\theta \in \Theta} \left\{ 
				||v(\theta) + \xi(\theta)||_\infty
			\right\} \\ 
			&\ge \min_{\theta \in \Theta} \{ ||v(\theta)||_\infty \} - \max_{\theta \in 
				\Theta} \{ ||\xi(\theta)||_\infty \} \\
			&\ge \min_{\theta \in \partial S} \{ ||v(\theta)||_\infty \} - \max_{\theta 
			\in \partial S} \{ ||\xi(\theta)||_\infty \}
		\end{align*}
		Given the initial condition $\theta_0$ to the perturbed model, the initial 
		frequency deviation is $\delta_0 = ||v(\theta_0) + \xi(\theta_0)||_\infty = 
		||\xi(\theta_0)||_\infty$, so applying \eqref{eq:perturb} and the lower bound on 
		$V^*(\partial S)$, we obtain
		\[
			\delta_0 = ||\xi(\theta)||_\infty < \min_{\theta \in \partial S} 
			||v(\theta)||_\infty - \max_{\theta \in \partial S} ||\xi(\theta)||_\infty 
			\le V^*(\partial S)
		\]
		Therefore $\gamma$ and $S$ satisfy the hypothesis of Theorem~\ref{thm:general} in 
		the perturbed model, and the theorem statements follow.
	\end{proof}

	Condition~\eqref{eq:perturb} bounds the scale of the perturbation $\xi(\theta)$. As we 
	will see, in many cases, the left-hand size of the equation is straightforward to 
	compute (or at least upper bound). The right-hand side of \eqref{eq:perturb} is 
	\textit{almost} identical to either $\widehat V(\gamma)$ or $\widehat V(\gamma, u_0)$ 
	(depending on whether $S$ is intersected with the winding cell); the only difference 
	is that the ``$D^{-1} f(\theta)$ is pointed outward from $S$'' constraint is removed.
	It is straightforward to obtain a lower bound on $\min_{\theta \in 
	\partial S} ||v(\theta)||_\infty$ with a minor relaxation to either $\widehat 
	V(\gamma)$ or $\widehat V(\gamma, u)$: 
	simply remove the $d^{-1}_{s(e)} f_{s(e)} - d_{t(e)}^{-1} f_{t(e)}$ constraints from 
	\eqref{prob2:ind-plus} and \eqref{prob2:ind-minus}. This lower bound can be used in 
	place of the left-hand side of \eqref{eq:perturb}. 
	
	In the IEEE 24-bus test case, we use random sampling to identify a point $\gamma 
	\in [\gamma_0, \gamma^*]$ for which $\min_{\theta \in \partial S} 
	||v(\theta)||_\infty \ge 0.0435$, with respect to the set $S = \Delta(\gamma) \cap 
	\Omega_{u_0}$, $u_0 = \vect 0_{11}$. The arc lengths in this particular $\gamma$ 
	range from 18.5 to 22.1 degrees, with a median of 20.3 degrees. Therefore, Theorem 
	\ref{thm:robust} guarantees that the IEEE 24-bus steady-state is robust to any 
	perturbations in parameters for which $||\xi(\theta_0)||_\infty + \max_{\theta \in 
	\partial S} ||\xi(\theta)||_\infty < 0.0435$.

	In the remaining subsections, we will apply Theorem~\ref{thm:robust} to particular 
	modes of disturbances: fluctuations in nominal power injections, changes in nodal 
	voltages, and changes in branch admittances. 
	
	\subsection{Perturbations of Nominal Injections}
	
	\begin{figure}
		\centering
		\includegraphics[clip, trim={1.5cm 0 0 0}, 
		width=0.8\linewidth]{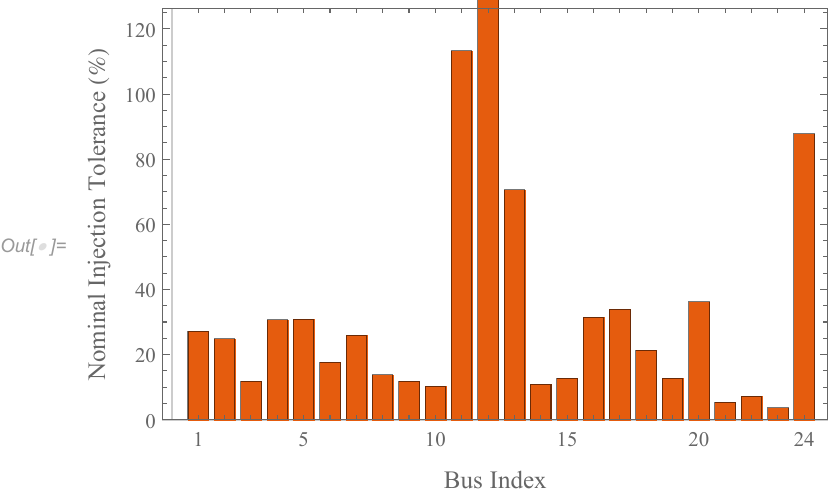}
		\caption{Relative tolerances of nominal power injections in the IEEE 24-bus 
		system. Each bar indicates the range of perturbations (as a percentage of the 
		nominal value) with respect to which the transient stability certificate still 
		holds. Note that these perturbations may occur simultaneously. Also note that bus 
		12 has a tolerance of 1500\% due to its small nominal injection.}
	 	\label{fig:pnom}
	\end{figure}
	
	Suppose that the perturbed model is identical to the original model, except the 
	vector of nominal frequency deviations has been shifted to $p^* + \Delta p^*$. It is 
	then clear from \eqref{def:freq-deviation} that the vector of frequency deviations 
	suffers the perturbation $\xi(\theta) = D^{-1} \Delta p^*$. This quantity is 
	constant, and condition~\eqref{eq:perturb} reduces to the condition 
	\[
		||D^{-1} \Delta p^*||_\infty < \frac 1 2 \min_{\theta \in \partial S} 
		||v(\theta)||_\infty
	\]
	for some $S = \Delta(\gamma) \cap \Omega_{u_0}$, with $\gamma \in [\gamma_0, 
	\gamma^*]$. In the IEEE 24-bus test case, a sufficient condition is $||\Delta 
	p^*||_\infty < 0.217~\text{pu}$. The median bus in this test case has a 
	nominal injection magnitude of $0.94~\text{pu}$, so our certificate guarantees that 
	the system is robust to disturbances of $23\%$ in the nominal 
	injection of this bus. Figure \ref{fig:pnom} plots the relative tolerance of all 
	buses in the system.
	
	\subsection{Perturbations of Voltage and Admittance Magnitudes}
	
	Next, we consider perturbations to nodal voltage magnitudes and branch admittance 
	magnitudes. Both of these values are encoded in the $\tilde a_{ij}$ and $a_{ij}$ 
	parameters, so these perturbations can be represented with perturbations $\Delta 
	\tilde a_{ij}$ and $\Delta a_{ij}$. The entries of the corresponding perturbation 
	vector are
	\[
		\xi_i(\theta) = d_i^{-1} \sum_{j \in \mathcal N(i)} \Delta \tilde a_{ij} + 
		\Delta a_{ij} \sin(\theta_i - \theta_j - \phi_{ij})
	\] 
	For simplicity, assume that $\Delta \tilde a_{ij} \le 0$ and $\Delta a_{ij} \le 0$ 
	(i.e., there is a loss in voltage magnitudes or branch admittances). 
	In order to compute $||\xi(\theta_0)||_\infty$, we first compute $\eta_{ij} = 
	\sin(\theta_i - \theta_j - \phi_{ij})$ using $\theta_0$, so that
	\[
		||\xi(\theta_0)||_\infty = \max_i \left\{ d_i^{-1} \left|\sum_{j \in \mathcal 
		N(i)} 
		\Delta \tilde a_{ij} + \eta_{ij} \Delta a_{ij} \right| \right\}
	\]
	Similarly, defining $\bar \eta_{ij} = \max\{ \sin( \gamma_{\{i, j\}} - \phi_{ij}), \; 
	\sin(\gamma_{\{i, j\}} + \phi_{ij}) \}$ as an upper bound on $|\sin(\theta_i - 
	\theta_j - \phi_{ij})|$ for $\theta \in \Delta(\gamma)$, we can bound
	\[
		\max_{\theta \in \partial S} ||\xi(\theta)||_\infty \le \max_{i} \left\{ 
		-d_i^{-1} \left( 
		\sum_{j \in \mathcal N(i)} \Delta \tilde a_{ij} + \bar \eta_{ij} \Delta a_{ij} 
		\right) \right\}
	\] 
	Then condition~\eqref{eq:perturb} is satisfied if the sum of these two quantities is less than 
	$\max_{\theta \in \partial S} ||v(\theta)||_\infty$. Note that this condition is a 
	set of linear constraints on $\Delta \tilde a_{ij}$ and $\Delta a_{ij}$.
	
	\begin{figure}
		\centering
		\includegraphics[clip, trim={1.5cm 0 0 0}, 
		width=0.8\linewidth]{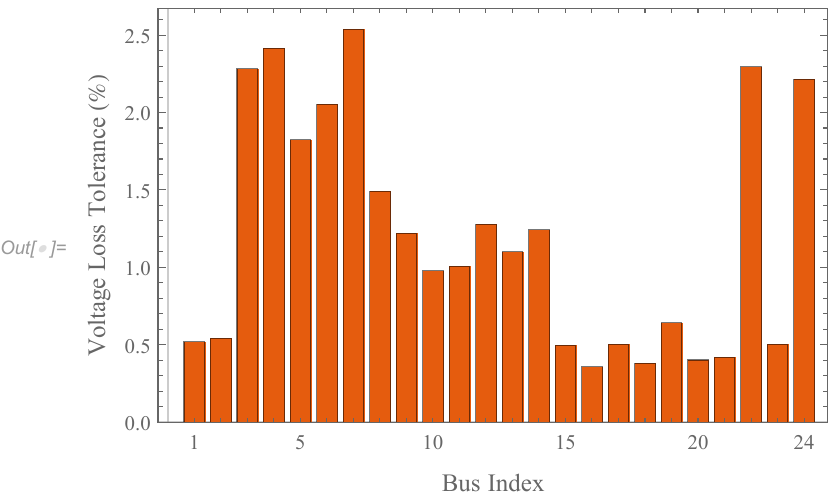}
		\caption{Tolerance for voltage loss in the IEEE 24-bus system, as a percentage of 
		the nominal voltage magnitude. If an individual 
		bus loses the fraction of voltage magnitude indicated by its corresponding bar in 
		the chart, then transient stability in the post-fault system is guaranteed.}
		\label{fig:volt}
	\end{figure}
	
	For a simple illustrative example, 
	suppose that one particular bus $\ell \in \mathcal V$ suffers a loss in voltage 
	magnitude, so that $E_\ell \to (1 - \alpha) E_\ell$ for some $\alpha \in (0, 1]$. 
	Then $\Delta \tilde a_{\ell j} = -\alpha(2 - \alpha) \tilde a_{\ell j}$, $\Delta 
	a_{\ell j} = -\alpha a_{\ell j}$, and $\Delta a_{j \ell} = -\alpha a_{j \ell}$ for 
	all $j \in 	\mathcal N(\ell)$, while the remaining perturbations are zero. In the 
	IEEE 24-bus	test case, we compute the largest value of $\alpha$ that satisfies the 
	previous equation for each $\ell \in \mathcal V$, using 0.0435 for the right-hand 
	side of the bound. These largest $\alpha$ are plotted in Figure \ref{fig:volt}. The 
	median bus can tolerate a 1\% loss of voltage magnitude before $\Delta a_{ij}$ and 
	$\Delta \tilde a_{ij}$ violate the above bound. This is much more restrictive than 
	the bound for nominal power injections, which is to be expected, given that we used 
	a conservative upper bound on $\max_{\theta \in \partial S} |\xi_i(\theta)|$ instead 
	of the exact value.  
}
	
	\section{Conclusion} 
	
	In this paper, we study transient stability in power networks consisting of
	droop-controlled inverters and frequency-dependent loads. We extend the
	notion of transient stability to include not only frequency synchronization
	but also operating constraints on nodal frequencies, angle differences,
	power injections, ramping, and storage reserves. To analyze the transients,
	we introduce a physically-meaningful Lyapunov-like function, and we
	re-cast the transient stability problem as an optimization problem that
	admits an efficient relaxation. \rev{We show that incorporating information from loop 
	flows (in the form of the winding vector) can make these transient stability 
	certificates less conservative. Finally, we show how these certificates can be used 
	to quantify the size of parameter disturbances to which the system is robust.}
	
	\rev{The model we use in this paper} is, of course, a highly simplified model for 
	frequency dynamics. Nonetheless, we hope that this paper provides a step toward
	understanding the fundamental behavior of future low-inertia power
	grids. Extensions of this work may offer rigorous answers to open
	theoretical questions about these systems. At what scale do the ubiquitous
	grid-following inverters harm system stability, and how can power engineers
	use droop-controlled inverters to mitigate this effect? How do legacy
	high-inertia generators affect transient behavior in power networks that
	are dominated by inverters? Future research may enrich this work with
	models of different types of generators, to better-understand frequency
	dynamics as power grids transition to low-inertia power sources.

	\appendices
 
	\section{Proof of Lemmas \ref{lem:probs1} and \ref{lem:probs2}}
	\label{appendix:prob}
\rev{ 
	
	\begin{proof}[Proof of Lemma \ref{lem:probs1}] 
		Let $\theta$ be the minimizing argument of Problem \ref{prob:minmax}, let $y_e = 
		d_{\rm cc}(\theta_i, \theta_j)$ be the counterclockwise angle difference across 
		each branch $e \in \mathcal E$, and define $f$, $\eta^+$, and $\eta^-$ according 
		to \eqref{prob2:f}--\eqref{prob2:eta-minus}. Because $\theta 
		\in \partial \Delta(\gamma)$, \eqref{prob2:y-bound} holds because $\theta \in 
		\cl(\Delta(\gamma))$, and there exists some edge $e \in \mathcal E$ such 
		that $y_e = \pm \gamma_e$. In the case where $y_e = \gamma_e$, let $z_e^+ = 1$, 
		so the two linear constraints in \eqref{prob2:ind-plus} activate. The first 
		constraint is just $y_e = \gamma_e$, which we have assumed is true. The second 
		constraint, $d_{s(e)}^{-1} f_{s(e)} - d_{t(e)}^{-1} f_{t(e)} \ge 0$, holds 
		because $D^{-1} f(\theta) = D^{-1} f$ points outward from $\Delta(\gamma)$. Thus 
		\eqref{prob2:ind-plus} is satisfied, and \eqref{prob2:ind-minus} and 
		\eqref{prob2:sos1} are satisfied by setting all other entries of $z^+$ and $z^-$ 
		equal to zero. In the case where $y_e = -\gamma_e$, we employ a complementary 
		argument with $z^-_e = 1$. In both cases, all constraints are satisfied, so $(f, 
		y, \eta^+, \eta^-, z^+, z^-)$ is feasible. Finally, the cost function 
		\eqref{prob2:cost} is equal to $V(\theta)$ (the cost function of Problem 
		\ref{prob:minmax}) when evaluated at $\theta$. Hence $\widehat 
		V(\gamma) \le V^*(\partial \Delta(\gamma))$. 
		
		To see that equality holds in the tree case, let us consider the argmin $(f, y, 
		\eta, z^+, z^-)$ of Problem \ref{prob:milp}. Because $G$ is a tree and $|y| \le 
		\gamma$, there always exists $\theta \in \Delta(\gamma)$ for which $y_e = d_{\rm 
		cc}(\theta_{s(e)}, \theta_{t(e)})$ for all $e \in \mathcal E$. Then 
		\eqref{prob2:y-bound}--\eqref{prob2:sos1} ensure that 
		$\theta \in \partial \Delta(\gamma)$, and 
		\eqref{prob2:cost}--\eqref{prob2:eta-minus} ensure that the cost 
		function of Problem \ref{prob:milp} is identical to $V(\theta)$. Hence 
		$V^*(\partial \Delta(\gamma)) \le \widehat V(\gamma)$ if $G$ is acyclic.
	\end{proof}

	The proof of Lemma \ref{lem:probs2} proceeds similarly, but it uses some 
	properties of the winding partition of the $n$-torus. Readers unfamiliar with the 
	winding partition are encouraged to read \ref{sect:winding-intro} and glance at 
	\cite{SJ-EYH-KDS-FB:18jv3}. The property that we need is the following lemma, which 
	shows how the winding partition relates to the boundaries of 
	phase-cohesive sets:
	
	\begin{lemma} \label{lem:topo}
		Let $\Omega_u$ be a winding cell, and let $\gamma \in (0, \pi \vect 1_m)$. 
		Consider the set $S = \Delta(\gamma) \cap \Omega_u$. Then $\partial S = \partial 
		\Delta(\gamma) \cap \Omega_u$.
	\end{lemma}

	\begin{proof}
		We first argue that $\cl(\Delta(\gamma)) \cap \partial \Omega_u = \emptyset$. If 
		$\theta \in \cl(\Delta(\gamma))$, then $|\theta_i - \theta_j| < \pi$ for all 
		$\{i, j\} \in \mathcal E$ (by the assumption that $\gamma < \pi \vect 1_m$). Then 
		there is a neighborhood around $\theta$ within which the winding numbers around 
		each cycle do not change, so $\theta$ belongs to the interior of a winding cell.
		
		First, using elementary properties of topology, we have 
		\[
			\partial S
			\subseteq \cl(S)
			\subseteq \cl(\Delta(\gamma)) \cap \cl(\Omega_u)
		\] 
		Because $\partial S \subseteq \cl(\Delta(\gamma))$, we have that $\partial S \cap 
		\partial \Omega_u = \emptyset$. Therefore, from the elementary property $\partial 
		S \subseteq \partial \Delta(\gamma) \cup \partial \Omega_u$, we obtain the result 
		$\partial S \subseteq \Delta(\gamma)$. Furthermore, because $\partial S \subseteq 
		\cl(\Omega_u)$ but $\partial S \cap \partial \Omega_u = \emptyset$, we have that 
		$\partial S \subseteq \interior(\Omega_u) \subseteq \Omega_u$. Hence 
		$\partial S \subseteq \partial \Delta(\gamma) \cap \Omega_u$.
		
		To show equality, we invoke the winding partition to write
		\begin{align*}
			\partial \Delta(\gamma) &= \partial \left( \bigcup_{v \in \Img(w_\Sigma)} 
			\Delta(\gamma) \cap \Omega_v \right) \\
			&\subseteq \bigcup_{v \in \Img(w_\Sigma)} \partial(\Delta(\gamma) \cap 
			\Omega_v)
		\end{align*}
		For $v \ne u$, the sets $\Omega_u$ and $\partial(\Delta(\gamma) \cap \Omega_v)$ 
		are disjoint, since we have shown that $\partial(\Delta(\gamma) \cap \Omega_v) 
		\subseteq \Omega_v$. Thus, intersecting both sides of the equation with 
		$\Omega_u$, we obtain $\partial \Delta(\gamma) \cap \Omega_u \subseteq 
		\partial(\Delta(\gamma) \cap \Omega_u) = \partial S$. This completes the proof.
	\end{proof}

	We now prove Lemma \ref{lem:probs2}.

	\begin{proof}[Proof of Lemma \ref{lem:probs2}] 
		The statement that $\widehat V(\gamma) \le \widehat V(\gamma, u)$ 
		is obvious, since Problem \ref{prob:milp} is a relaxation of Problem 
		\ref{prob:exact}.
				
		To show that $\widehat V(\gamma, u) \le 
		V^*(\partial S)$, let $\theta$ be the minimizing argument of Problem 
		\ref{prob:minmax}. As in the proof of Lemma \ref{lem:probs1}, select the values 
		of the decision variables $(f, y, \eta, z^+, z^-)$ accordingly to satisfy 
		\eqref{prob2:f}--\eqref{prob2:sos1}, thereby ensuring that the cost function 
		\eqref{prob2:cost} is equal to $V(\theta)$. Because $\theta \in \Omega_u$, 
		\cite[Theorem 3.5]{SJ-EYH-KDS-FB:18jv3} guarantees the existence of $x \in 
		\vect 1_n^\top$ such that $y = B^\top x + 2\pi C_\Sigma^\dagger u$. Multiplying 
		across by $C_\Sigma$, we obtain
		\[
			C_\Sigma y = C_\Sigma B^\top x + 2\pi C_\Sigma
			C_\Sigma^\dagger u = 2\pi u
		\]
		We have performed two simplifications in this equation. First, 
		$\Img(C_\Sigma^\top)$ is the cycle space, which is identical to $\ker(B)$ 
		\cite[Theorem 9.5]{FB:20}, so the $C_\Sigma B^\top x$ term vanishes. Second, the 
		summation structure in \eqref{eq:winding-num} implies that $u \in 
		\Img(C_\Sigma)$, so the orthogonal projection matrix $C_\Sigma C_\Sigma^\dagger$ 
		has no effect on $u$. Thus $(f, y, \eta)$ satisfies all of the constraints of 
		Problem \ref{prob:exact}, so $\widehat V(\gamma, u) \le V^*(\partial S)$. 
		
		Next, we will show that $V^*(\partial S) \le \widehat V(\gamma, 
		u)$. Let $(f, y, \eta, z^+, z^-)$ be the minimizing argument of Problem 
		\ref{prob:exact}. 
		Consider the equation $y = B^\top x + 2\pi C_\Sigma^\dagger u$. Note that 
		$\Img(B^\top) = (\Img(C_\Sigma^\dagger))^\bot$, so we can decompose $y = y_1 + 
		y_2$ with $y_1 \in \Img(B^\top)$ and $y_2 = \Img(C_\Sigma^\dagger)$, and the 
		equation can be split into $y_1 = B^\top x$ and $y_2 = 2\pi C_\Sigma^\dagger u$. 
		The first equation has a unique solution $x \in \vect 1_n^\bot$, while the second 
		equation is true because $C_\Sigma y = 2\pi u$, so there is a unique point $x \in 
		\vect 1_n^\bot$ that satisfies $y = B^\top x + 2\pi C_\Sigma^\dagger u$. It 
		follows from \cite[Theorem 3.5]{SJ-EYH-KDS-FB:18jv3} that there exists $\theta 
		\in 
		\Omega_u$ such that $y_e = d_{\rm cc}(\theta_{s(e)}, \theta_{t(e)})$ for all $e 
		\in \mathcal E$. From the remaining constraints in 
		Problem \ref{prob:exact}, we can see that $\theta \in \partial \Delta(\gamma) 
		\cap \Omega_u$, so it follows from Lemma \ref{lem:topo} that $\theta \in \partial 
		S$. Furthermore, the constraints imply that the velocity vector is 
		pointed outward. Thus $\theta$ is within the feasible set of Problem 
		\ref{prob:minmax}, and the identical values of the cost functions imply that 
		$V^*(\partial S) \le \widehat V(\gamma, u)$.  
	\end{proof}

}

	\ifCLASSOPTIONcaptionsoff
	\newpage
	\fi
	
	\bibliographystyle{plainurl+isbn}
	\bibliography{alias,Main,FB,New}

\begin{thebibliography}{10}

\bibitem{NA-SG:13}
N.~Ainsworth and S.~Grijalva.
\newblock A structure-preserving model and sufficient condition for frequency
  synchronization of lossless droop inverter-based {AC} networks.
\newblock {\em IEEE Transactions on Power Systems}, 28(4):4310--4319, 2013.
\newblock \href {http://dx.doi.org/10.1109/TPWRS.2013.2257887}
  {\path{doi:10.1109/TPWRS.2013.2257887}}.

\bibitem{TA-RP-SV:79}
T.~Athay, R.~Podmore, and S.~Virmani.
\newblock A practical method for the direct analysis of transient stability.
\newblock {\em IEEE Transactions on Power Apparatus and Systems},
  98(2):573--584, 1979.
\newblock \href {http://dx.doi.org/10.1109/TPAS.1979.31940}
  {\path{doi:10.1109/TPAS.1979.31940}}.

\bibitem{FB-SM:15}
F.~Blanchini and S.~Miani.
\newblock {\em Set-Theoretic Methods in Control}.
\newblock Springer, 2015, ISBN 9783319179322.

\bibitem{PB-AL-AT-SW:15}
P.~Bonami, A.~Lodi, A.~Tramontani, and S.~Wiese.
\newblock On mathematical programming with indicator constraints.
\newblock {\em Mathematical Programming}, 151:191--223, 2015.
\newblock \href {http://dx.doi.org/10.1007/s10107-015-0891-4}
  {\path{doi:10.1007/s10107-015-0891-4}}.

\bibitem{JCB-TC-LD:18}
J.~C. Bronski, T.~Carty, and L.~DeVille.
\newblock Configurational stability for the {Kuramoto}\textendash{}{Sakaguchi}
  model.
\newblock {\em Chaos: An Interdisciplinary Journal of Nonlinear Science},
  28(10):103109, 2018.
\newblock \href {http://dx.doi.org/10.1063/1.5029397}
  {\path{doi:10.1063/1.5029397}}.

\bibitem{FB:20}
F.~Bullo.
\newblock {\em Lectures on Network Systems}.
\newblock Kindle Direct Publishing, {1.4} edition, July 2020, ISBN
  978-1986425643.
\newblock With contributions by J. Cort{\'e}s, F. D\"orfler, and S.
  Mart{\'\i}nez.
\newblock URL: \url{http://motion.me.ucsb.edu/book-lns}.

\bibitem{MCC-DMD-RA:93}
M.~C. Chandorkar, D.~M. Divan, and R.~Adapa.
\newblock Control of parallel connected inverters in standalone {AC} supply
  systems.
\newblock {\em IEEE Transactions on Industry Applications}, 29(1):136--143,
  1993.
\newblock \href {http://dx.doi.org/10.1109/28.195899}
  {\path{doi:10.1109/28.195899}}.

\bibitem{HDC-CCC-GC:95}
H.-D. Chiang, C.~C. Chu, and G.~Cauley.
\newblock Direct stability analysis of electric power systems using energy
  functions: {T}heory, applications, and perspective.
\newblock {\em Proceedings of the IEEE}, 83(11):1497--1529, 1995.
\newblock \href {http://dx.doi.org/10.1109/5.481632}
  {\path{doi:10.1109/5.481632}}.

\bibitem{TC-RD-IA-PJ:16}
T.~Coletta, R.~Delabays, I.~Adagideli, and P.~Jacquod.
\newblock Topologically protected loop flows in high voltage {AC} power grids.
\newblock {\em New Journal of Physics}, 18(10):103042, 2016.
\newblock \href {http://dx.doi.org/10.1088/1367-2630/18/10/103042}
  {\path{doi:10.1088/1367-2630/18/10/103042}}.

\bibitem{FDS-DA:07}
F.~{De~Smet} and D.~Aeyels.
\newblock Partial entrainment in the finite {Kuramoto}--{Sakaguchi} model.
\newblock {\em Physica D: Nonlinear Phenomena}, 234(2):81--89, 2007.
\newblock \href {http://dx.doi.org/10.1016/j.physd.2007.06.025}
  {\path{doi:10.1016/j.physd.2007.06.025}}.

\bibitem{FD-FB:09z}
F.~D{\"o}rfler and F.~Bullo.
\newblock Synchronization and transient stability in power networks and
  non-uniform {K}uramoto oscillators.
\newblock {\em SIAM Journal on Control and Optimization}, 50(3):1616--1642,
  2012.
\newblock \href {http://dx.doi.org/10.1137/110851584}
  {\path{doi:10.1137/110851584}}.

\bibitem{CG-PW-PA-RA-MB-RB-QC-CF-SH-SK-WL-RM-DP-NR-DR-AS-MS-CS:99}
C.~Grigg et~al.
\newblock {The IEEE Reliability Test System-1996. A report prepared by the
  Reliability Test System Task Force of the Application of Probability Methods
  Subcommittee}.
\newblock {\em IEEE Transactions on Power Systems}, 14(3):1010--1020, 1999.
\newblock \href {http://dx.doi.org/10.1109/59.780914}
  {\path{doi:10.1109/59.780914}}.

\bibitem{LLG:12}
L.~L. Grigsby, editor.
\newblock {\em Power System Stability and Control}.
\newblock CRC Press, 3 edition, 2012, ISBN 9781439883204.

\bibitem{DG-JSB-FD:19}
D.~Gro{\ss}, J.~S. Brouillon, and F.~D{\"o}rfler.
\newblock The effect of transmission-line dynamics on grid-forming dispatchable
  virtual oscillator control.
\newblock {\em IEEE Transactions on Control of Network Systems},
  6(3):1148--1160, 2019.
\newblock \href {http://dx.doi.org/10.1109/TCNS.2019.2921347}
  {\path{doi:10.1109/TCNS.2019.2921347}}.

\bibitem{NERC-04}
NERC~Steering Group.
\newblock {Technical Analysis of the August 14, 2003, Blackout: What Happened,
  Why, and What Did We Learn?}
\newblock Technical report, North American Electric Reliability Council,
  Princeton Forrestal Village, Princeton, NJ, USA, July 2004.

\bibitem{JMG-JCV-JM-LGDV-MC:11}
J.~M. Guerrero, J.~C. Vasquez, J.~Matas, L.~G. de~Vicuna, and M.~Castilla.
\newblock Hierarchical control of droop-controlled {AC} and {DC} microgrids--a
  general approach toward standardization.
\newblock {\em IEEE Transactions on Industrial Electronics}, 58(1):158--172,
  2011.
\newblock \href {http://dx.doi.org/10.1109/TIE.2010.2066534}
  {\path{doi:10.1109/TIE.2010.2066534}}.

\bibitem{SJ-EYH-KDS-FB:18jv3}
S.~Jafarpour, E.~Y. Huang, K.~D. Smith, and F.~Bullo.
\newblock Flow and elastic networks on the $n$-torus: {Geometry,} analysis and
  computation.
\newblock {\em SIAM Review}, 2019.
\newblock Submitted.
\newblock URL: \url{https://arxiv.org/pdf/1901.11189v3.pdf}.

\bibitem{NJ-AK:03}
N.~Janssens and A.~Kamagate.
\newblock Loop flows in a ring {AC} power system.
\newblock {\em International Journal of Electrical Power \& Energy Systems},
  25(8):591--597, 2003.
\newblock \href {http://dx.doi.org/10.1016/S0142-0615(03)00017-6}
  {\path{doi:10.1016/S0142-0615(03)00017-6}}.

\bibitem{SK-WD-SPN-FT-KS:19}
S.~Kundu, W.~Du, S.~P. Nandanoori, F.~Tuffner, and K.~Schneider.
\newblock Identifying parameter space for robust stability in nonlinear
  networks: A microgrid application.
\newblock In {\em {A}merican {C}ontrol {C}onference}, pages 3111--3116, July
  2019.
\newblock \href {http://dx.doi.org/10.23919/ACC.2019.8814324}
  {\path{doi:10.23919/ACC.2019.8814324}}.

\bibitem{SK-SPN-KK-SG-IAH:19}
S.~Kundu, S.~P. Nandanoori, K.~Kalsi, S.~Geng, and I.~A. Hiskens.
\newblock Distributed barrier certificates for safe operation of inverter-based
  microgrids.
\newblock In {\em {A}merican {C}ontrol {C}onference}, pages 1042--1047, July
  2019.
\newblock \href {http://dx.doi.org/10.23919/ACC.2019.8815296}
  {\path{doi:10.23919/ACC.2019.8815296}}.

\bibitem{PK:94}
P.~Kundur.
\newblock {\em Power System Stability and Control}.
\newblock McGraw-Hill, 1994, ISBN 007035958X.

\bibitem{HS-YK:86}
H.~Sakaguchi and Y.~Kuramoto.
\newblock A soluble active rotater model showing phase transitions via mutual
  entertainment.
\newblock {\em Progress of Theoretical Physics}, 76(3):576--581, 1986.
\newblock \href {http://dx.doi.org/10.1143/PTP.76.576}
  {\path{doi:10.1143/PTP.76.576}}.

\bibitem{JWSP-FD-FB:12u}
J.~W. Simpson-Porco, F.~D{\"o}rfler, and F.~Bullo.
\newblock Synchronization and power sharing for droop-controlled inverters in
  islanded microgrids.
\newblock {\em Automatica}, 49(9):2603--2611, 2013.
\newblock \href {http://dx.doi.org/10.1016/j.automatica.2013.05.018}
  {\path{doi:10.1016/j.automatica.2013.05.018}}.

\bibitem{PV-FFW-RLC:85}
P.~Varaiya, F.~F. Wu, and R.-L. Chen.
\newblock Direct methods for transient stability analysis of power systems:
  {R}ecent results.
\newblock {\em Proceedings of the IEEE}, 73(12):1703--1715, 1985.
\newblock \href {http://dx.doi.org/10.1109/PROC.1985.13366}
  {\path{doi:10.1109/PROC.1985.13366}}.

\bibitem{TLV-KT:16}
T.~L. Vu and K.~Turitsyn.
\newblock Lyapunov functions family approach to transient stability assessment.
\newblock {\em IEEE Transactions on Power Systems}, 31(2):1269--1277, 2016.
\newblock \href {http://dx.doi.org/10.1109/TPWRS.2015.2425885}
  {\path{doi:10.1109/TPWRS.2015.2425885}}.

\bibitem{TLV-KT:17}
T.~L. Vu and K.~Turitsyn.
\newblock A framework for robust assessment of power grid stability and
  resiliency.
\newblock {\em IEEE Transactions on Automatic Control}, 62(3):1165--1177, 2017.
\newblock \href {http://dx.doi.org/10.1109/TAC.2016.2579743}
  {\path{doi:10.1109/TAC.2016.2579743}}.

\bibitem{KX-HXL-CS-JHVS:20}
K.~{Xi}, H.~X. {Lin}, C.~{Shen}, and J.~H. {Van Schuppen}.
\newblock Multilevel power-imbalance allocation control for secondary frequency
  control of power systems.
\newblock {\em IEEE Transactions on Automatic Control}, 65(7):2913--2928, 2020.
\newblock \href {http://dx.doi.org/10.1109/TAC.2019.2934014}
  {\path{doi:10.1109/TAC.2019.2934014}}.

\bibitem{MKSY-SHS:99}
M.~K.~S. Yeung and S.~H. Strogatz.
\newblock Time delay in the {Kuramoto} model of coupled oscillators.
\newblock {\em Physical Review Letters}, 82(3):648, 1999.
\newblock \href {http://dx.doi.org/10.1103/PhysRevLett.82.648}
  {\path{doi:10.1103/PhysRevLett.82.648}}.

\bibitem{LZ-DH:18}
L.~Zhu and D.~Hill.
\newblock Stability analysis of power systems: A network synchronization
  perspective.
\newblock {\em SIAM Journal on Control and Optimization}, 56(3):1640--1664,
  2018.
\newblock \href {http://dx.doi.org/10.1137/17M1118646}
  {\path{doi:10.1137/17M1118646}}.

\end{thebibliography}
	
	\begin{IEEEbiography}[{\includegraphics[height=1.3in]{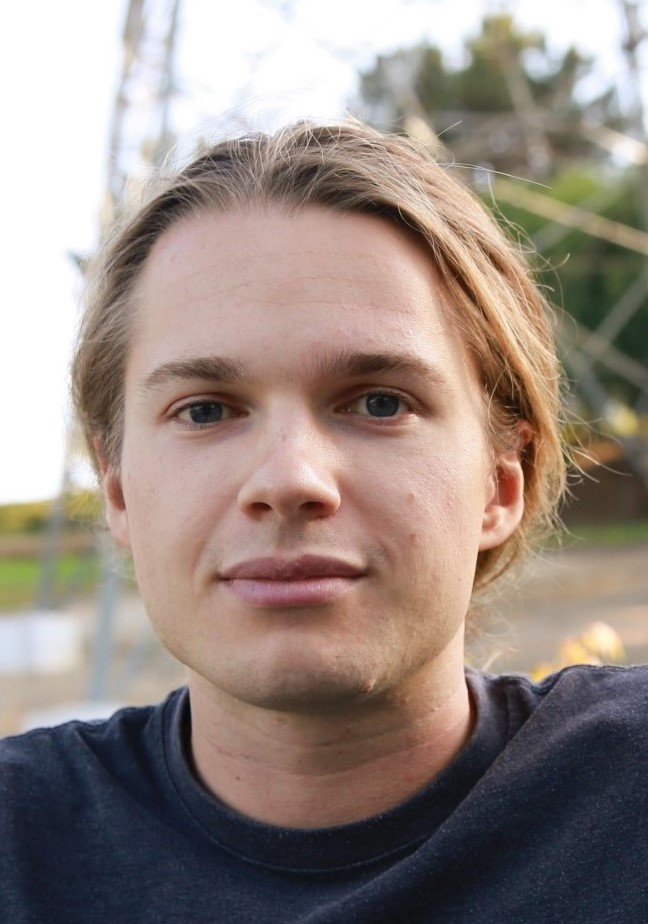}}]{Kevin
			D. Smith} is a Ph.D. student with the Center for Control, Dynamical Systems 
			and Computation at the University of California, Santa Barbara. He holds a 
			B.S. in physics from Harvey Mudd College and an M.S. from the Department of 
			Electrical and Computer Engineering at UCSB. He is interested in dynamics, 
			control, and identification of network systems, particularly infrastructure 
			networks. 
	\end{IEEEbiography}
	
	\begin{IEEEbiography}[{\includegraphics[height=1.3in]{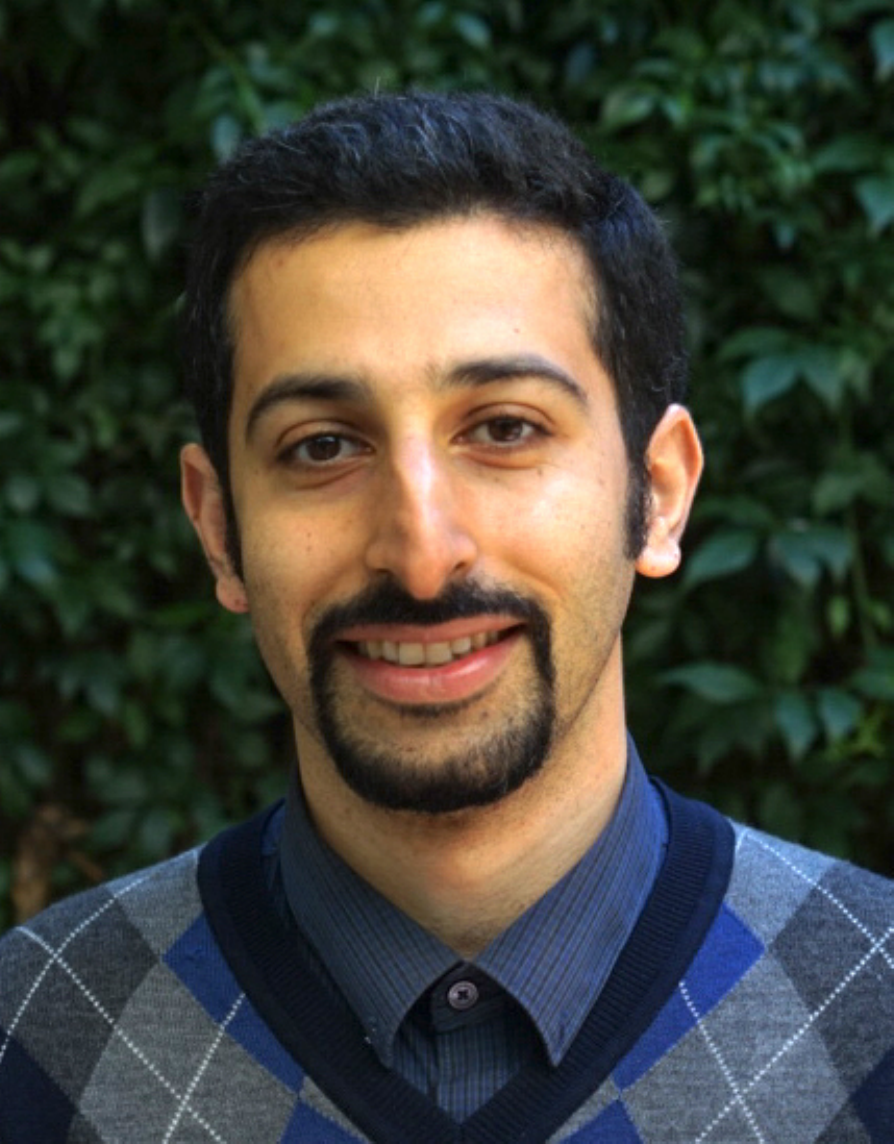}}]{Saber Jafarpour}(M'16)
		is a Postdoctoral researcher with the Mechanical Engineering Department and
		the Center for Control, Dynamical Systems and Computation at the University
		of California, Santa Barbara. He received his Ph.D. in 2016 from the
		Department of Mathematics and Statistics at Queen's University. His
		research interests include analysis of network systems with application to
		power grids and geometric control theory.
	\end{IEEEbiography}
	
	\begin{IEEEbiography}[{\includegraphics[height=1.3in]{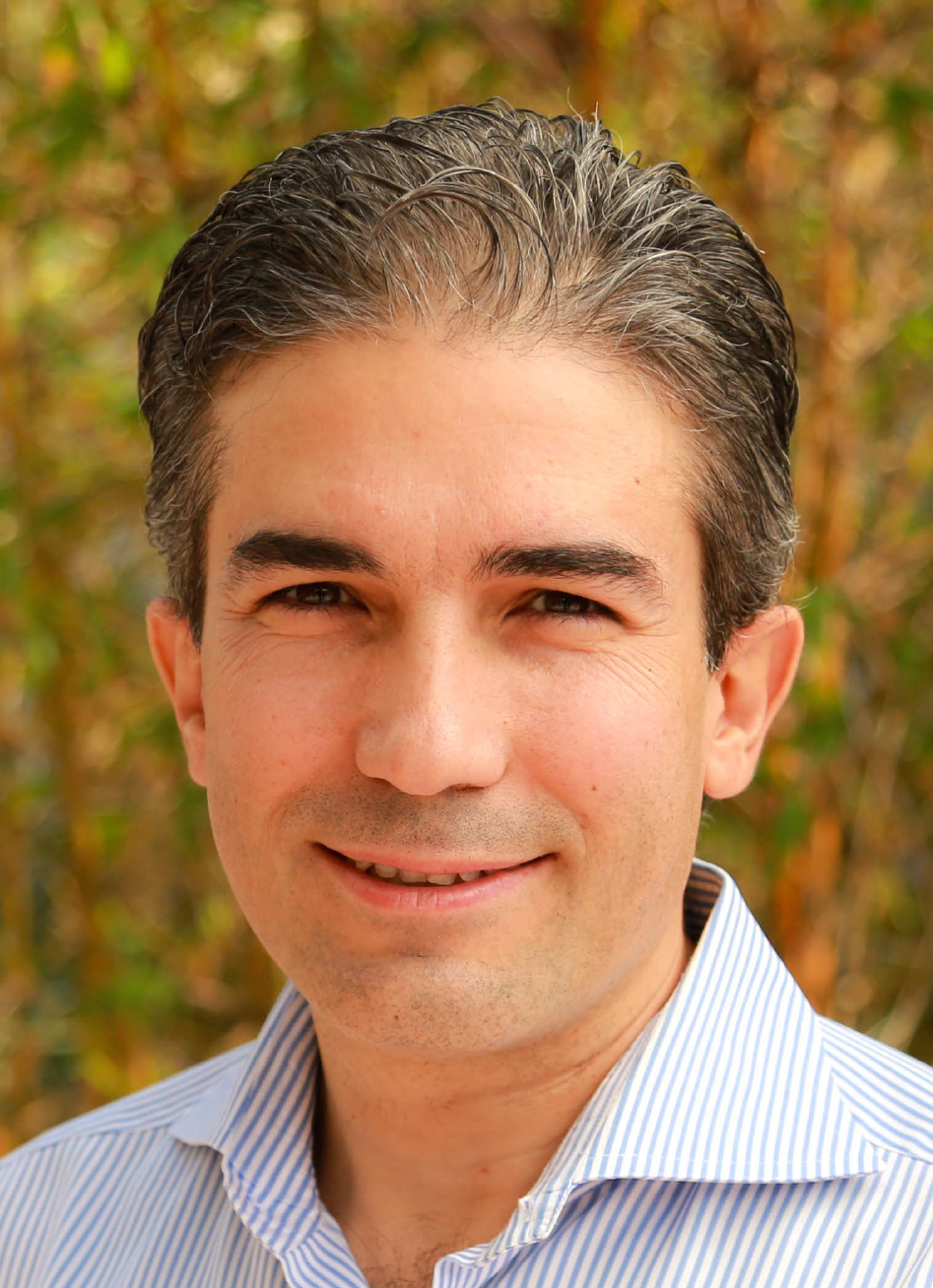}}]{Francesco Bullo}(S'95-M'99-SM'03-F'10)
 is a Professor with the Mechanical Engineering Department and the Center
 for Control, Dynamical Systems and Computation at the University of
 California, Santa Barbara. He was previously associated with the
 University of Padova (Laurea degree in Electrical Engineering, 1994), the
 California Institute of Technology (Ph.D. degree in Control and Dynamical
 Systems, 1999), and the University of Illinois. He served on the editorial
 boards of IEEE, SIAM, and ESAIM journals and as IEEE CSS President. His
 research interests focus on network systems and distributed control with
 application to robotic coordination, power grids and social networks. He
 is the coauthor of “Geometric Control of Mechanical Systems” (Springer,
 2004), “Distributed Control of Robotic Networks” (Princeton, 2009), and
 “Lectures on Network Systems” (Kindle Direct Publishing, 2020, v1.4).  He
 is a Fellow of IEEE, IFAC, and SIAM.
	\end{IEEEbiography}
	
\end{document}